\documentclass[fleqn,usenatbib]{mnras}

\usepackage{newtxtext,newtxmath}

\usepackage[T1]{fontenc}

\DeclareRobustCommand{\VAN}[3]{#2}
\let\VANthebibliography\thebibliography
\def\thebibliography{\DeclareRobustCommand{\VAN}[3]{##3}\VANthebibliography}

\usepackage{graphicx}	
\usepackage{amsmath}	

\graphicspath{ {./images/} }
\usepackage[caption=false]{subfig}
\usepackage{booktabs}
\usepackage{siunitx}

\title[MWA EoR1 Upper limits]{Epoch of Reionization Power Spectrum Limits from Murchison Widefield Array Data Targeted at EoR1 Field}

\author[M. Rahimi et al.]{
M. Rahimi,$^{1,2}$\thanks{E-mail: rahimi@student.unimelb.edu.au}
B.~Pindor,$^{1,2}$,
J. L. B.~Line,$^{3,2}$,
N.~Barry,$^{1,2}$,
C.~M.~Trott,$^{3,2}$,
R.~L.~Webster,$^{1,2}$,
C.~H.~Jordan,$^{3,2}$
\newauthor
M.~Wilensky,$^{4}$,
S.~Yoshiura,$^{1,5}$,
A.~Beardsley,$^{6}$,
J.~Bowman,$^{7}$,
R.~Byrne,$^{4}$,
A.~Chokshi,$^{1,2}$,
B.~J.~Hazelton,$^{4,8}$
\newauthor
K.~Hasegawa,$^{9}$,
E.~Howard,$^{10}$,
B.~Greig,$^{1,2}$,
D.~Jacobs,$^{7}$,
R.~Joseph,$^{3,2}$,
M.~Kolopanis,$^{7}$,
C.~Lynch,$^{3,2}$
\newauthor
B.~McKinley,$^{3,2}$,
D.~A.~Mitchell,$^{10,2}$,
S.~Murray,$^{7}$,
M.~F.~Morales,$^{4}$,
J.~C.~Pober,$^{11}$,
K.~Takahashi,$^{12}$
\newauthor
S.~J.~Tingay,$^{3,2}$
R.~B.~Wayth,$^{3,2}$,
J.~S.~B.~Wyithe,$^{1,2}$,
Q.~Zheng,$^{13}$
\\
\\
$^{1}${The University of Melbourne, School of Physics, Parkville, VIC 3010, Australia}\\
$^{2}${ARC Centre of Excellence for All Sky Astrophysics in 3 Dimensions (ASTRO-3D)}\\
$^{3}${International Centre for Radio Astronomy Research, Curtin University, Perth, WA 6845, Australia}\\
$^{4}${University of Washington, Department of Physics, Seattle, WA 98195, USA}\\
$^{5}${Mizusawa VLBI Observatory, National Astronomical Observatory Japan, 2-21-1 Osawa, Mitaka, Tokyo 181-8588, Japan}\\
$^{6}${Winona State University, Department of Physics, Winona, MN 55987, USA}\\
$^{7}${Arizona State University, School of Earth and Space Exploration, Tempe, AZ 85287, USA} \\
$^{8}${University of Washington, eScience Institute, Seattle, WA 98195, USA}\\
$^{9}${Graduate School of Science, Nagoya University, Japan}\\
$^{10}${CSIRO Astronomy and Space Science (CASS), PO Box 76, Epping, NSW 1710, Australia}\\
$^{11}${Brown University, Department of Physics, Providence, RI 02912, USA}\\
$^{12}${Faculty of Science, Kumamoto University, 2-39-1 Kurokami, Kumamoto 860-8555, Japan}\\
$^{13}${Shanghai Astronomical Observatory, China}\\
}

\date{Accepted XXX. Received YYY; in original form ZZZ}

\pubyear{2021}

\begin{document}
\label{firstpage}
\pagerange{\pageref{firstpage}--\pageref{lastpage}}
\maketitle

\begin{abstract}
Current attempts to measure the 21cm Power Spectrum of neutral hydrogen during the Epoch of Reionization are limited by systematics which produce measured upper limits above both the thermal noise and the expected cosmological signal. These systematics arise from a combination of observational, instrumental, and analysis effects. In order to further understand and mitigate these effects, it is instructive to explore different aspects of existing datasets. One such aspect is the choice of observing field. To date, MWA EoR observations have largely focused on the EoR0 field. In this work, we present a new detailed analysis of the EoR1 field. The EoR1 field is one of the coldest regions of the Southern radio sky, but contains the very bright radio galaxy Fornax-A. The presence of this bright extended source in the primary beam of the interferometer makes the calibration and analysis of EoR1 particularly challenging. We demonstrate the effectiveness of a recently developed shapelet model of Fornax-A in improving the results from this field. We also describe and apply a series of data quality metrics which identify and remove systematically contaminated data.
With substantially improved source models, upgraded  analysis algorithms and enhanced data quality metrics, we determine EoR power spectrum upper limits based on analysis of the best $\sim$14-hours data observed during 2015 and 2014 at redshifts 6.5, 6.8 and 7.1, with the lowest $2\sigma$ upper limit at z=6.5 of $\Delta^2 \leq (73.78 ~\mathrm{mK)^2}$ at $k=0.13~\mathrm{h~ Mpc^{-1}}$, improving on previous EoR1 measurement results.
 
\end{abstract}




\section{Introduction}
Over cosmic time, hydrogen has gone through three major transitions: ionized hydrogen post-nucleosynthesis in the primordial plasma, neutral hydrogen as the universe cooled by expansion, and finally reionized hydrogen as a result of the formation of the first stars and galaxies \citep{Furlanetto:2006, Morales_Wyithe:2010,  Pritchard:2012}.
Our goal is to measure the H$\rm{I}$ spin temperature and neutral fraction via its 21-cm signal to follow the evolution of neutral gas in the IGM during the Epoch of Reionization (EoR).

The GMRT (Giant Metrewave Radio Telescope, \citet{Paciga:2013}), LOFAR
(LOw Frequency Array\footnote{http://www.lofar.org}, \citet{Yatawatta:2013}), PAPER (Precision Array for Probing the Epoch of Reionization\footnote{http://eor.berkeley.edu}, \citet{Parsons:2010}), LWA (Long Wavelength Array\footnote{http"//lwa.unm.edu}, \citet{Ellingson:2009}) and the MWA (Murchison Widefield Array \footnote{http://www.mwatelescope.org}, \citet{Bowman:2013}, \citet{Tingay:2013}) are first generation instruments with the primary goal of detecting the EoR signal.

In reality, detection of such a faint signal ($\sim 10~\mathrm{mK}$) is complicated; i.e. foregrounds, which are radio emission from galactic and extragalactic sources, are up to five orders of magnitude brighter than the EoR signal \citep{Jelic2008}. They inherently have smooth spectra as opposed to the cosmological 21-cm signal which varies rapidly in frequency. Therefore, by developing suitable analysis approaches, they can be separated from the EoR signal in power space\citep{Dillon:2014,Dillon:2015b,Trott:2016}. The other major challenge is the systematic errors in calibration due to instrumental effects such as beam models \citep{Joseph2020} or spectral behaviour of electric components\citep{trott_wayth_2016}, imperfect sky models\citep{Barry:2016,byrne2019,Zhang2020}, radio frequency interference (RFI) \citep{Offringa:2015} and ionospheric distortions\citep{Jordan2017,Trott2018,degasperin2018}. In addition, due to low SNR, we need to integrate over long observation times to average down the noise. Over the last years, many efforts have been taken to develop novel analysis methodologies to overcome these challenges and detect the signal\citep{morales2019}

Due to the low sensitivity of first generation telescopes, the only effective measurement of the H\rm{I} signal is via statistical measurements like the power spectrum. Therefore, we quantify the statistics of the brightness temperature contrast between the 21-cm signal and the CMB and take advantage of the isotropy and homogeneity of the universe to average the 3D k-space power spectrum over spherical shells of constant spatial scale $\lvert \boldsymbol{k} \rvert \mathrm{Mpc^{-1}}$. The current experiments placed upper limits on the EoR power spectrum, such as MWA\citep{Dillon:2014,Dillon:2015,Beardsley2016,Barry2019,Li2019,Trott2020}, GMRT\citep{Paciga:2013}, LOFAR\citep{Patil2017, Mertens2020} and PAPER\citep{Kolopanis2019}. Due to high contamination by systematics, they have not detected the signal yet. However, in addition to learning invaluable lessons in design and analysis, the results are getting informative enough to constrain the astrophysical models\citep{greig2021a,greig2021b}.

The next generation of instruments, such as SKA-Low (Square Kilometer Array, \citet{Koopmans2014}), HERA (Hydrogen Epoch of Reionization Array, \citet{DeBoer:2016}) and NenuFAR (New Extension in Nan\c{c}y Upgrading LOFAR, \citet{zarka:hal-01196457}), are under design/designed with substantially larger collecting area and hence higher sensitivity with respect to their first generation counterparts. Therefore, beyond detection on limited redshift or spatial scales, they will be able to characterise the EoR power spectrum with more details and also potentially map the ionized bubbles of hydrogen during the EoR.

Efforts to calculate upper limits for the cosmological H\rm{I} power spectra using first generation of instruments, with on-going efforts to improve the precision and quality of the algorithms and their implementation \citep{Paciga:2013, Dillon:2015, Beardsley2016, Patil2017, Li2019, Barry2019}, have produced the most recent $2\sigma$ upper limit of $ 1.8 \times 10^3~\mathrm{mK^2}$ at $k=0.14~\mathrm{h~Mpc^{-1}}$ and $z=6.5$ from 110 hours of MWA field EoR0 high-band data \citep{Trott2020}. 

The very bright ($\sim 100s~\mathrm{Jy}$) and extended radio galaxy Fornax-A is located in the EoR1 field. Therefore, this field requires a more detailed sky model than the EoR0 field. \citet{Trott2020} measured the limits on the power spectrum of deep observations over three MWA EoR fields, and found that EoR1 produced higher limits than the other two fields at each studied redshift. In the present work, $\sim$14 hours of Phase-I MWA data, targeting the EoR1 field and observed during 2014 and 2015, is analyzed using the \textit{RTS/CHIPS} (Real Time System, \citet{Mitchell:2008,Ord:2010}/ The Cosmological H$\rm{I}$ Power Spectrum,\citet{Trott:2016}) pipeline to measure the upper limits at three redshifts: 6.5, 6.8 and 7.1.
The calibration procedure including the source models were improved relative to the previous work \citep{Trott2020} and customised to obtain more precise calibration solutions. In addition, the application of well established data quality metrics such as window power \citep{Beardsley2016}, removed contaminated data from the analysis dataset. By applying new metrics, new systematic signatures in data were examined. This provided a multi-layered understanding of data anomalies and hence allowed the development of well-targeted approaches for systematic mitigation, either by resolving shortcomings in the analysis or by more effective data selection methods. The measured upper limits over 14 hours of data improve on the recent limits from EoR1 published at \citet{Trott2020}.

The rest of this paper is organized as follows. In Section 2 we describe the instrument and the observational data we processed in this study and our initial understanding of the EoR1 field. In section 3 we introduce the analysis pipeline and sky model that were applied to the data. Section 4 compares the datasets observed in 2015/2014, targeted at the EoR1 field, with the corresponding data observed in 2013 or targeted at the EoR0 field. We also compare data variations across different nights and pointings, and finally present the results. Lastly, in section 5 the analysis results and future possible analysis improvements are discussed.
\section{Instrument and Observations}
\subsection{The Murchison Widefield Array} 
The Murchison Widefield Array is a radio interferometer located at the Murchison Radio Observatory,  a radio-quiet site in Western Australia. While supporting several science programs, it was primarily built with the goal of detecting the EoR signal. The technical design of MWA and its science capabilities are discussed in \citet{Tingay:2013} and \citet{Bowman:2013} respectively. MWA has been later upgraded in several phases \citep{wayth2018,beardsley2019}.

The MWA Phase-\rm{I} configuration, which was used to obtain the data in this work, consisted of 128 antenna-tiles each comprising 16 dual polarization dipoles placed on a regular grid. These tiles were distributed within a radius of 1.5 km, with a dense core of 50m for the purpose of EoR experiments.  Each dipole is sensitive to the entire sky and is optimized to operate in the $80-300~\mathrm{MHz}$ frequency band, out of which a processing bandwidth of $30.72~\mathrm{MHz}$ is selected. The radio signals from the dipoles of a tile are combined in an analogue beamformer. For EoR experiments, the beamformer steers the targeted direction of observation in steps of $6.8^{\circ}$ ,each taking 27 minutes. The primary beam for each direction, or "pointing", provides a wide field-of-view of about $25^{\circ}$ at $150~\mathrm{MHz}$. In this work, we analyse the five central pointings, including two pointings before and after the zenith transit, inclusive.  They are labeled in Local Sidereal Time order as Minus2, Minus1, Zenith, Plus1 and Plus2, respectively. 
The radio signal is then transmitted to receivers in which they are digitized and filtered into 24 coarse channels of $1.28~\mathrm{MHz}$ width, the edges of which need to be flagged prior to our analysis to avoid the aliasing effect introduced by channelization. Then they are passed to the correlator. The correlator then cross-multiplies the signals between all pairs of antennas at 10kHz resolution.These cross correlations are called visibilities, which are averaged over frequency and time and then transmitted to be archived. For the data included here, the correlation output resolution is 2s in time and 40kHz in frequency. The averaged data are written as snapshots of cadence of 112 seconds, named as one single \textit{observation}.
\subsection{EoR Observing Program}
\begin{figure*}
\centering
\subfloat[MWA EoR fields]{\includegraphics[width=0.95\linewidth]{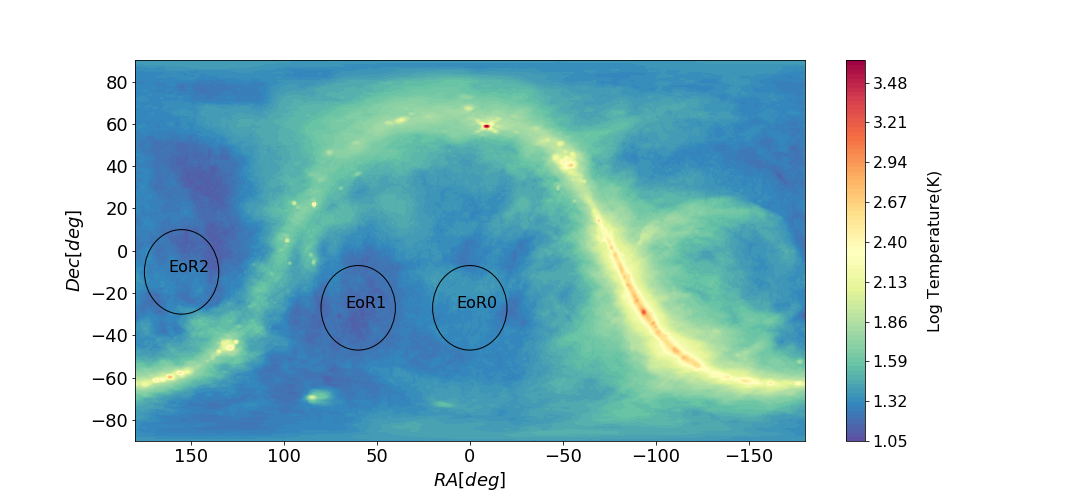}
\label{fig:mwa-eor-fields}
}
\\
\subfloat[MWA EoR1 field centered at ($4^h, -27^{\circ}$), including Fornax-A at ($3.3^h, -37.1^{\circ}$)]{\includegraphics[width=0.45\textwidth]{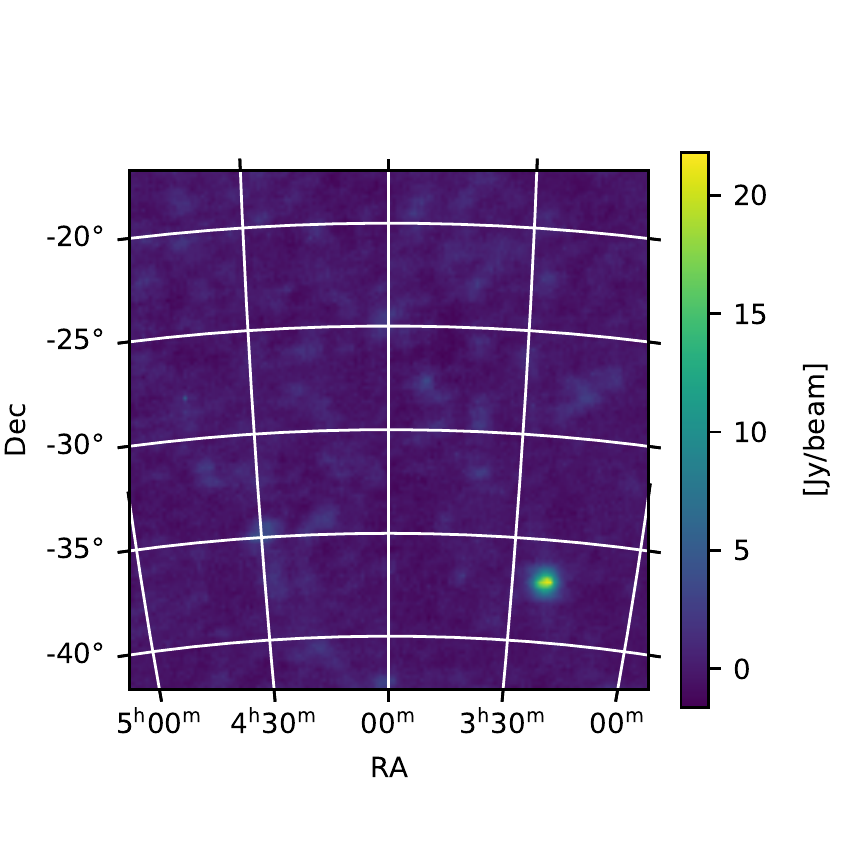}
\label{fig:eor1-field}}
\subfloat[Fornax-A Phase-I/II image.]{\includegraphics[width=0.45\textwidth]{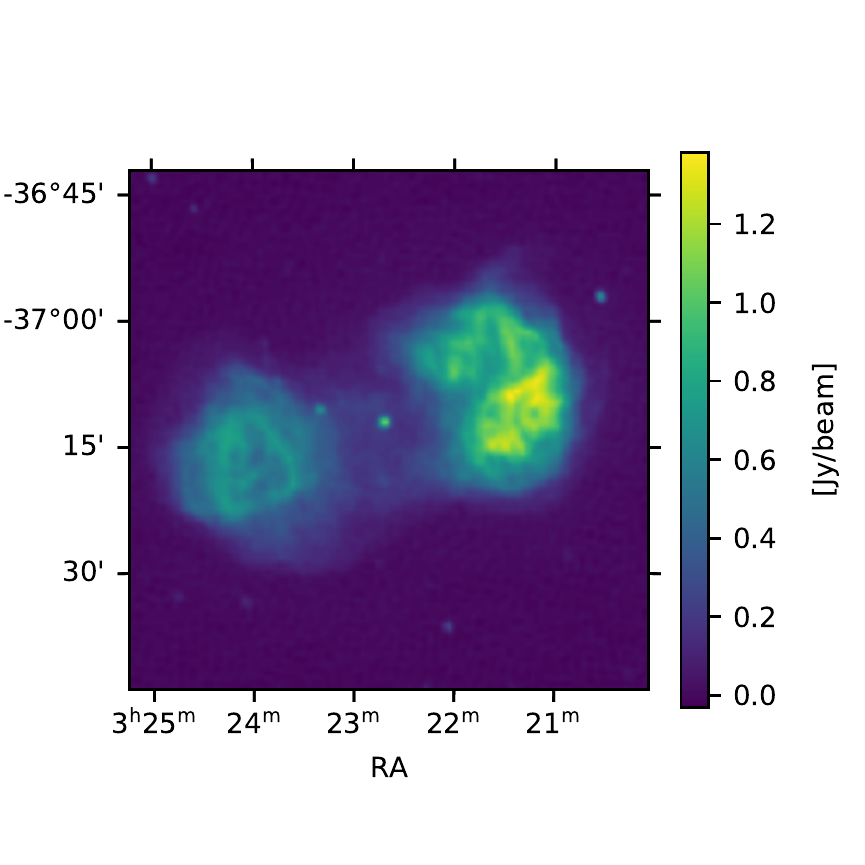}\label{fig:fornax}}
\caption{a)MWA EoR fields, b) EoR1 field and c) Fornax-A image.}
\end{figure*}

MWA EoR observations are collected over three frequency bands: High ($167 - 197~\mathrm{MHz}$) centred at redshift $z\approx6.8$, Low ($139 - 167~\mathrm{MHz}$) centered at $z\approx8.2$ and Ultra-low ($75 - 100~\mathrm{MHz}$) centered at $z\approx16$. There are three targeted fields for EoR observations which are relatively devoid of galactic emission and extragalactic bright sources, named EoR0, EoR1 and EoR2, as shown in figure \ref{fig:mwa-eor-fields}. Their coordinates are presented in Table \ref{tab:fields}. 

\begin{table}
    \centering
    \begin{tabular}{lll}
    \toprule
    & \multicolumn{2}{c}{centered at} \\
    Field & RA & Dec \\
    \midrule
    EoR0 & $0^{h}$ & $-27^{\circ}$ \\
    EoR1 & $4^{h}$ & $-27^{\circ}$ \\
    EoR2 & $10.33^{h}$  &  $-10^{\circ}$ \\
    \bottomrule
    \end{tabular}
    \caption{MWA EoR fields. EoR0 and EoR1 transit over zenith at the MWA latitude.}
    \label{tab:fields}
\end{table}
The physical characteristics of the field of interest for this work, EoR1, and its comparison to the widely studied one, EoR0, are discussed in Section \ref{sec:comparison}. With MWA phase-I configuration, thousands of hours of EoR data were collected from mid-2013 to mid-2016.

\subsection{Data}\label{sec_data}
This work aims to analyze and measure the power spectrum limit of the EoR signal from the EoR1 field, observed in High band by the MWA Phase-I configuration. \citet{Trott2020} compared the power spectra from integrations of 300, 500, 680 and 820 EoR1 observations. They showed that adding data beyond 300 observations (until $\sim$ 800 observations) did not improve the power spectrum, due to the saturation by systematics (For more discussion refer to section \ref{sec:pipeline}). Therefore, partially motivated by finite computing resources, we aimed to demonstrate improvements on a final dataset of 400-500 observations. Initially, we calibrated a pilot dataset spanning all of the observation nights of 2015 (September to December). Based on the results of that processing, we excluded from further analysis nights with a relatively high rate of calibration failure, a high ionospheric metric, a high number of dead tiles or with data format issues. We also included the zenith pointing observations from 2014 (spanning August to September). Therefore, we began our fiducial analysis with 767 successfully calibrated snapshots from 2015 and 2014. This dataset is then refined as described in section \ref{sec:data-metrics}. The final upper limits presented in section \ref{sec:limits} are obtained from 459 selected observations ($\sim$ 14-hour integration), covering five central pointings (including the zenith). There are 171 ($\sim 37\%$ of total) zenith observations in the final set.
\subsection{Target Field of EoR1}\label{sec:eor1}
The targeted fields for EoR studies with MWA have been chosen to be in relatively cold parts of sky, but the presence of some bright radio sources such as Fornax-A in EoR1, shown in figure \ref{fig:eor1-field}, is unavoidable. The bright and complex radio emission of Fornax-A (total flux density of $\sim 500~\mathrm{Jy}$ at $189~\mathrm{MHz}$\citep{bernardi2013}), shown in figure \ref{fig:fornax}, makes the process of calibration and foreground removal more complicated. Previous MWA studies of this field include  \citet{McKinley2014} who modeled the spectral energy distribution of Fornax-A and \citet{Procopio2017} who improved the source catalogue in the EoR1 field by combining both GLEAM and GMRT $150~\mathrm{MHz}$ data. Since Fornax-A is an extended source with spatial and spectral structure, it is essential to develop a model which effectively represents its complicated morphology. This is further discussed in \ref{sec:catalogue}.

In contrast, the EoR0 field does not have anything as bright or extended as Fornax-A, and thus has been used more in MWA EoR analysis \citep{Dillon:2015,Beardsley2016,Barry2019,Li2019,Trott2020}. We compare and contrast analysis of the two fields in section \ref{sec:comparison}.


\section{Data Calibration and Power Spectrum Calculation}\label{sec:pipeline}
\citet{Beardsley:2013} showed that a full season of MWA observation on two fields, yielding an integration time of $900 + 700~ \mathrm{hour}$, can potentially detect the EoR signal with a S/N of 14 on the amplitude. Were the measurements only noise-dominated, the larger integrations would average down the uncorrelated noise and lead to detection. However, in practice, the systematic errors, originated both from astrophysical and non-astrophysical sources, largely contribute to the power, exceeding the thermal noise level. There have been many efforts in the community to understand about different systematics and the effective strategies to mitigate them; e.g. the residual power due to imperfect calibration models \citep{Offringa:2015,Patil2017,Procopio2017,morales2018,Li2019} and the spectral contamination from faint unmodeled sources \citep{Barry:2016}, excess power due to imperfect telescope beam models \citep{Beardsley2016,Li2019,Barry2019,Joseph2020}, the contamination due to RFI \citep{Offringa:2015} and following possible contamination due to RFI excision \citep{Offringa2019}, the introduced spectral structure due to instrumental chromaticity of interferometers \citep{trott_wayth_2016} and the power bias due to ionospheric distortion effects \citep{Jordan2017,Trott2018}. \citet{Trott2020} found that integrating beyond 300 observations (up to 820) did not significantly improve their results, implying the existence of a systematic floor of less than $\sim26~\mathrm{hours}$.

While systematic contamination can be mitigated through different techniques employed in analysis pipelines; it is also possible to employ data quality metrics which detect and exclude highly contaminated data as described in section \ref{sec:results}.

\begin{figure}
    \centering
    \includegraphics[width=\columnwidth]{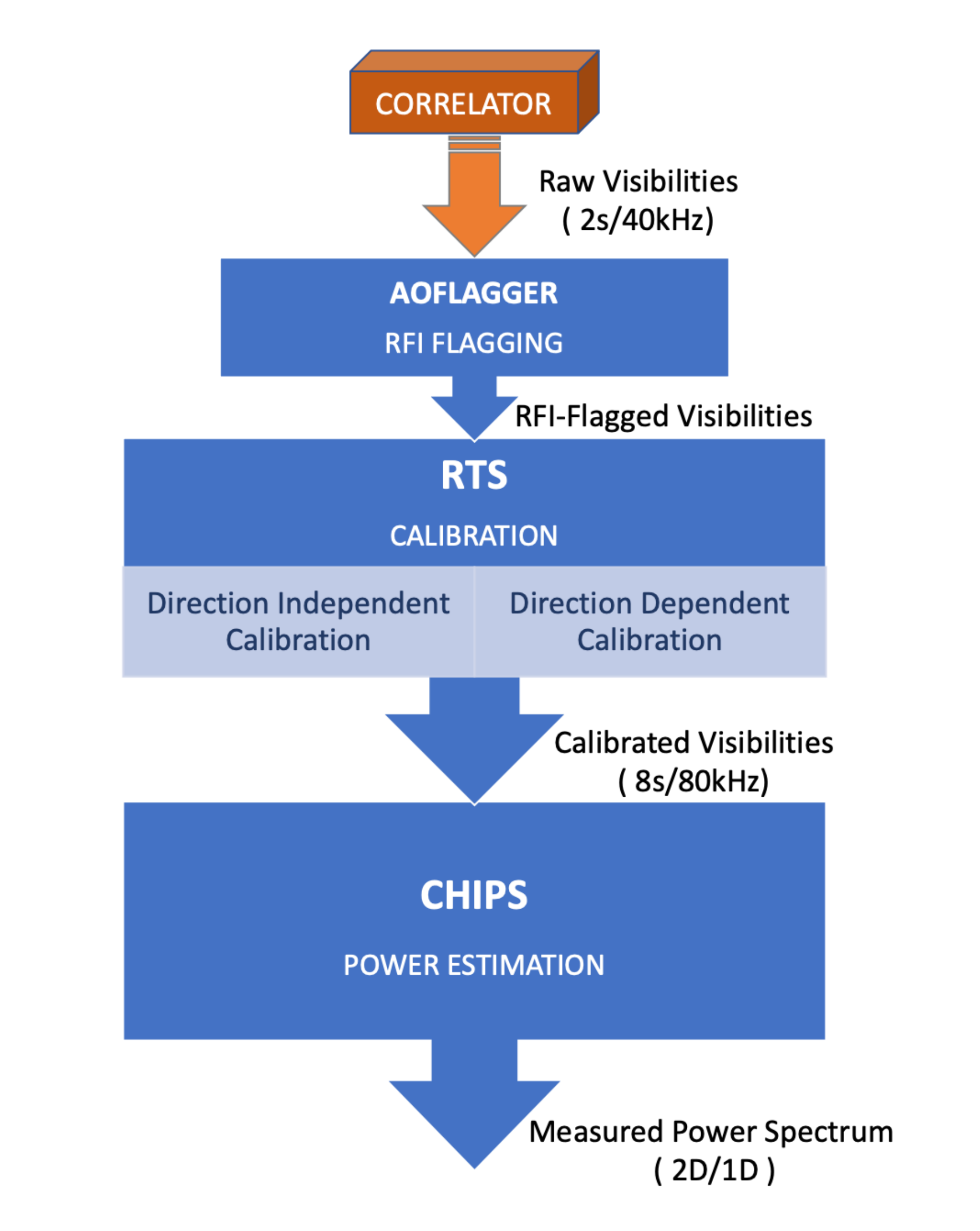}
    \caption{The data flow from correlator to software pipeline: The visibilities are passed to AOFLAGGER to be RFI-flagged, then to RTS to be calibrated and then to CHIPS to measure the power spectrum.}
    \label{fig:pipe}
\end{figure}

Analysis of a dataset to measure the power spectrum of the EoR signal requires three significant processes: (i) flagging of data affected by RFI, poor observing conditions or other anomalies, (ii) calibration and (iii) calculation of the power spectrum.
We implement the RTS/CHIPS processing pipeline to analyze the data in this work \citep{Jacobs:2016}. In this section, the current state of the pipeline is described, schematically shown in the diagram \ref{fig:pipe}.
\subsection{RFI Flagging}\label{sec:preprocess}
It is essential to mitigate the effect of RFI by detecting the contaminated data prior to processing. AOFLAGGER\footnote{http://aoflagger.sourceforge.net}  \citet{Offringa2010,Offringa2012} is the primary algorithm applied to detect and flag the RFI-contaminated data. In addition, the algorithm also flags five 40kHz channels in each coarse band: two 40kHz channels on either side of the band to avoid the aliasing effect introduced by the channelizing procedure, and the middle 40kHz band which has a DC offset \citep{Offringa:2015}. Practically, the RFI occupancy for most channels is either very low ($<5\%$) or very high ($>90\%$) in case of contaminated data. Therefore, at the end, we fully flag any channel which contains more than 80\% RFI occupancy to conservatively exclude any highly contaminated data.

In addition, as will be discussed in section \ref{sec:metrics}, we apply the technique of Sky-Subtracted Incoherent Noise Spectra (SSINS, \citet{Wilensky2019}), to mitigate the ultra-faint RFI contamination.
SSINS estimates the noise by taking advantage of the negligible sky variance over short time intervals of the interferometer measurements; and hence subtracts two time-adjacent visibility measurements and then averages them for all the baselines in the observation. This tool provides more sensitivity to interference, such as digital TV (DTV), hidden below the thermal noise level of individual baselines.

\subsection{RTS Calibration Pipeline}

The data is calibrated using the Real Time System (RTS, \citet{Mitchell:2008,Ord:2010}). The RTS was initially developed for real-time calibration and imaging of MWA data, but is now employed as an offline pipeline.

The RTS has a number of features aimed specifically at dealing with large low-frequency data including: measurement and correction of ionospheric effects, direction-dependant beam models, baseline-dependant averaging, use of point and extended source models, as well as parallelisation over frequency.

The RTS may be used in one of two modes, producing either calibrated residual visibilities or images. In this work, the first mode is used to create calibrated visibilities, which will be added together on a uvw-grid in CHIPS to form the power spectrum. 

A detailed description of the algorithm is discussed in \citet{Mitchell:2008}, which is reviewed here.
The RTS accomplishes the calibration process through two major steps;
\begin{description}
\item[Step 1:]
For each observation, a compound calibrator source is created, made up of the 1000 apparently brightest sources available in the source catalogue. The source catalogue is created by cross-matching with PUMA \citep{Line2017}. Full details of PUMA and this cross-matched catalogue can be found in \citet{Line2020}, but we include a brief description in section \ref{sec:catalogue}. The single compound calibrator provides a model for the visibilities. By fitting the observed visibilities to this model, one can determine the direction-independent complex voltage gain or \textit{Jones matrices} for each antenna.
\item[Step 2:] Bright sources, in comparison to faint sources, are less affected by flux contamination via the point-spread function of the remaining sources in the field, and thus should be peeled first to reduce subsequent error. Thus, all sources are ranked in order of apparent brightness. Then the model visibilities of the 1000 brightest sources are subtracted from the visibility set and then each source is individually peeled/calibrated in a Calibration Measurement Loop. The Calibration Measurement Loop performs these steps for each source individually in subsequent order: a) the model visibilities for the source are added back to the visibility set and the phase centre of the observation is rotated toward the actual position of the source; b) the gain and positional offset of the source due to ionospheric refractive effects are measured and corrected; c) the full direction-dependent calibration is computed and applied, but only for the five strongest sources due to computational limitations. For the rest of the  sources, only ionospheric corrections are computed and applied; d) once the measurements converge, they are used to subtract the source from the visibility set, thus improving the model visibility iteratively. This step is performed on an 8s cadence. Averaging the visibilities to 8s resolution increases the SNR while provides the required time resolution for sampling ionospheric fluctuations which is less than a minute. Once the calibrated residual visibilities are created, they are averaged to 80kHz in frequency for power estimation. This again provides higher SNR while resolves in the expected spectral fluctuations of EoR.
\end{description}

\subsection{Source Catalogue and Extended Source Modelling}\label{sec:catalogue}
Three sky models, which we label puma2017a, puma2017b, and puma2020, are used in this work to measure the effect of improved sky models on the power spectrum limits. All three used the GLEAM catalogue \citep{Hurley-Walker2017,Wayth:2015} as a base, cross-matched using PUMA to other catalogues to provide additional positional and spectral information (for details see \citet{Line2020}). The first sky model, puma2017a, is based on the work of \citet{Procopio2017}, which used TGSS GMRT~\citep{Intema2017} images to create higher resolution source models within $\sim10^\circ$ of the EoR1 field centre. The GLEAM catalogue was still being finalised at the time of this publication. Once the GLEAM catalogue had been finalised, the extended models from \citet{Procopio2017} were retained, and all other sources were replaced with updated models based on the full release version of GLEAM. We label this model puma2017b. RTS has the capability of digesting models of extended sources such as shapelet-based ones. A shapelet-model for Fornax-A is included in the sky model which is used in both calibration and subtraction. For both puma2017 sky models, the same model of Fornax-A was used. This was a shapelet model, based on phase I MWA data, as detailed in~\citet{Riding2017}. 

The puma2020 sky model includes an updated Fornax-A shaplet model, which utilizes a combination of MWA Phase I and Phase II data to increase angular resolution. For details on this model, please see \citet{Line2020}. In addition, an in-situ adjustment was made to the flux density and spectral index of Fornax-A in order to better match the observed flux density and spectral index in a sample of zenith data.

The EoR1 limits published in~\citet{Trott2020} used previously calibrated and peeled data products that were gradually processed over two years, using whichever sky model was available at time of processing. A re-processing of that entire data set would require a large investment of computational resources. We estimate 50\% of the data in the~\citet{Trott2020} EoR1 limit used the puma2017a model, and 50\% the puma2017b model.

All of the observations used in the  calculation of final limits in this paper are processed with the puma2020 sky model.
\subsection{Data Metrics}\label{sec:data-metrics}
The RTS calibration process also provides a number of metrics which we use to evaluate data quality and determine the suitability of each processed observation for power spectrum estimation. The three most important metrics are:
\begin{itemize}
    \item Calibration-Based Metrics: RTS provides the amplitude and phase of complex gain solutions for each tile. These solutions are a useful tool to evaluate the calibration quality since they reflect how smoothly the calibration procedure is performed over the frequency for each tile, both in shape and amplitude. In addition we can diagnose any tile with non consistent profile relative to others.
    \item IonoQA: The state of the ionosphere during the observation can be measured as a single value for ionospheric activity  by combining the median source offset and amount of offset directionality \citep{Jordan2017}. This "IonoQA" metric value varies from 2 (low ionospheric activity) to multiple 10s (high ionospheric activity). \citet{Trott2018} demonstrated the need to exclude  ionospherically active data to avoid bias in cosmological power spectrum measurements.
    \item Visibility Noise RMS: Since calibrated visibilities are averaged to 8-$\mathrm{s}$ resolution, each of 112-$\mathrm{s}$ observations consists of 14 measurement sets or equivalently 7 pairs of opposing even/odd cadences\footnote{Each 112-$\mathrm{s}$ snapshot (observation) accommodates a  series of 14 measurement sets, each covering 8-$\mathrm{s}$ integration time. We alternately label the 8-$\mathrm{s}$ cadences as \textit{even} and \textit{odd}. Therefore, there are 7 pairs of opposing even/odd cadences in each observation.}. Visibility Noise RMS is calculated as the standard deviation of differences between two sets of visibilities in each pair and is denoted as $\sigma_{vis}$. The reported RMS values throughout this work are calculated while the first and last time-samples are excluded due to some extra known errors,  introduced by partial flagging or missing data samples. In addition three 80kHz channels at the beginning, centre and end of each coarse band are excluded once calculating the RMS, to avoid the discontinuity due to the effect of either aliasing or DC offsets. The related frequency gaps in the x-axis of RMS plots, presented in section \ref{sec:results}, are also removed.
\end{itemize}

\subsection{Power Measurement - CHIPS}
In this work, CHIPS (the Cosmological HI Power Spectrum Estimator, \citet{Trott:2016}) is used to measure the power spectrum. It is an inverse variance estimator which computes the power from gridded, calibrated visibilities.

As mentioned before, we are attempting to study the physical properties of the hydrogen gas during EoR by probing the statistics of its temperature field, which are affected by the reionization process. By denoting temperature fluctuation (relative to the mean) as $T(\boldsymbol{r})$ at vector position $\boldsymbol{r}$, we represent the auto-correlation function of the temperature fluctuations signal as $\xi(\boldsymbol{r})$:
\begin{equation}
    \xi(\boldsymbol{r}) = \big< T(\boldsymbol{r_0})
    T(\boldsymbol{r_0} + \boldsymbol{r}) \big>_{\boldsymbol{r_0}}
\end{equation}
in which $<.>$ denotes the averaging over the sample volume, $Vol$. The Gaussian statistics of a signal is well described by its power spectrum, $P(\boldsymbol{k})$, in k-space. Under the assumption of the isotropic 21-cm signal, it provides the signal power spectrum over different spatial scales of k ($=|\boldsymbol{k}|$). 

The power spectrum is defined as the Fourier Transform of the auto-correlation function:
\begin{equation}
    P(k) = \int_{Vol} \xi(\boldsymbol{r})exp(-2\pi i \boldsymbol{k}.\boldsymbol{r}) d\boldsymbol{r}
\end{equation}
By replacing the auto-correlation function, the power spectrum equals to the spatial covariance of the signal, integrated over the spatial volume $Vol$:
\begin{equation}
    P(k) = \frac{1}{Vol}\big<\Tilde{T}(\boldsymbol{k})\Tilde{T}^*(\boldsymbol{k})\big>
\end{equation}
On the other hand, the interferometer measures the sky brightness as visibilities in ${u,v,w}$ domain. Given the flat-sky approximation, the measured visibilities, V(u,v) are the double Fourier Transform of the sky brightness, I(l,m). This can be represented as:
\begin{equation}
    V(u,v) = \int \int_{source} I(l,m) e^{-2\pi i (ul + vm )}dl dm
\end{equation}
where $u,v$ are Fourier
modes of the measured visibility and $l,m$ are the corresponding angular scales, which are represented by modes perpendicular to the line-or-sight, or the angular modes,$k_\perp$, in k-space. Additionally, the modes parallel to the line-of-sight, $k_{\parallel}$, may be mapped with spectral channels of measurement, while $k^2 = k_{\perp}^2 + k_{\parallel}^2$. The sky brightness is linearly related to brightness temperature at radio regime and, hence, the Power Spectrum can be obtained as:
\begin{equation}
    P(k) = \frac{1}{Vol}\big<\Tilde{V}(\boldsymbol{k})\Tilde{V}^*(\boldsymbol{k})\big>
\end{equation}
CHIPS starts off by performing the coherent integration of calibrated visibilities via gridding them all on to uv-plane per w-snapshot. This procedure creates the data cube in $(u,v,f)$. Next, each cell is Fourier Transformed along frequency and finally squared.

Measured power spectrum and upper limits are reported as the 1D power spectrum which is defined as:
\begin{equation}
    \Delta^2(k) \equiv \frac{k^3}{2\pi^2}P(k)
\end{equation}
and is the integrated total power over spatial scale $k$ in $\mathrm{mK^2}$.

\subsection{Power Spectrum Averaging Scheme}
\begin{figure}
    \centering
    \includegraphics[width=0.7\columnwidth]{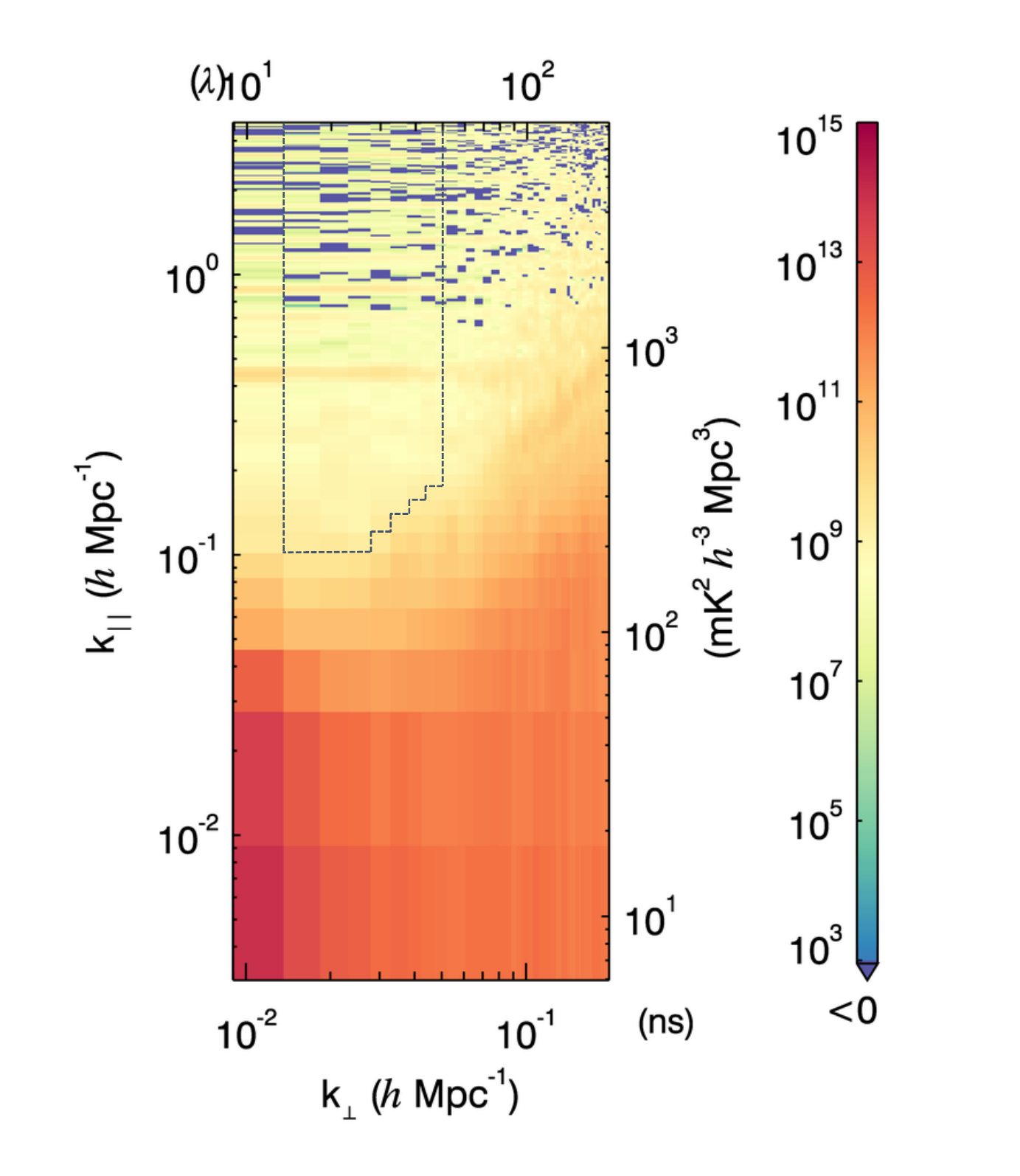}
    \caption{A typical 2D power spectrum for an integration of 20 observations. The region that is marked with the dashed line shows the selected range of ($k_{\perp},k_{\parallel}$) of which the measurements are included in averaging from 3D power spectrum to 1D.}
    \label{fig:k-selection}
\end{figure}

While the spherically-averaged 1D power spectrum (over 3D spatial scales of $k=|\boldsymbol{k}|$) provides the required cosmological measurements, the 2D cylindrically-averaged power spectrum is an informative tool to learn about the power at different regions of 2D space. Figure \ref{fig:k-selection} shows an example 2D power spectrum. It provides the power distribution over angular ($k_\perp$) and line of sight ($k_{\parallel}$) modes and will be used throughout this work for diagnostics. Since foregrounds vary smoothly as a function of frequency, they contaminate the low $k_{\parallel}$ modes. However, the instrument chromaticity couples the foregrounds into higher $k_{\parallel}$ modes which shapes a characteristic wedge in 2D space, called \textit{foreground wedge}. We can perform our measurement in the region above foreground wedge which is called the \textit{EoR window}. 

As discussed in \citet{Barry2019}, \citet{Li2019} and \citet{Trott2020}, cuts in k-space need to be implemented when averaging to the 1D power spectrum to exclude regions contaminated by foreground and instrumental systematics.  Specifically these are:
\begin{itemize}
    \item $12\lambda > k_{\perp} > 50\lambda$; to avoid contaminated low $k_{\perp}$ and poor uv-coverage at high $k_{\perp}$.
    \item $k_{\parallel} > 0.1~\mathrm{h Mpc^{-1}}$; to avoid the still present leakage from the foreground wedge at low $k_{\perp}$.
    \item $k_{\parallel} > 3.5 k_{\perp}$; to shift the horizon limit slightly higher preventing sub-horizon leakage. 
\end{itemize}
The selected k-space region is marked by a dashed line in figure \ref{fig:k-selection}. The same cuts and binning scheme are applied throughout the paper to average the 3D k-space to 1D.
\subsection{Improvements to Calibration Pipeline}\label{sec:improvement}
In this section, we demonstrate the major changes to the calibration pipeline, including those of the source catalogues and the peeling algorithm, and their effect in 1D power space. 
\begin{figure}
    \centering
    \includegraphics[width=\columnwidth]{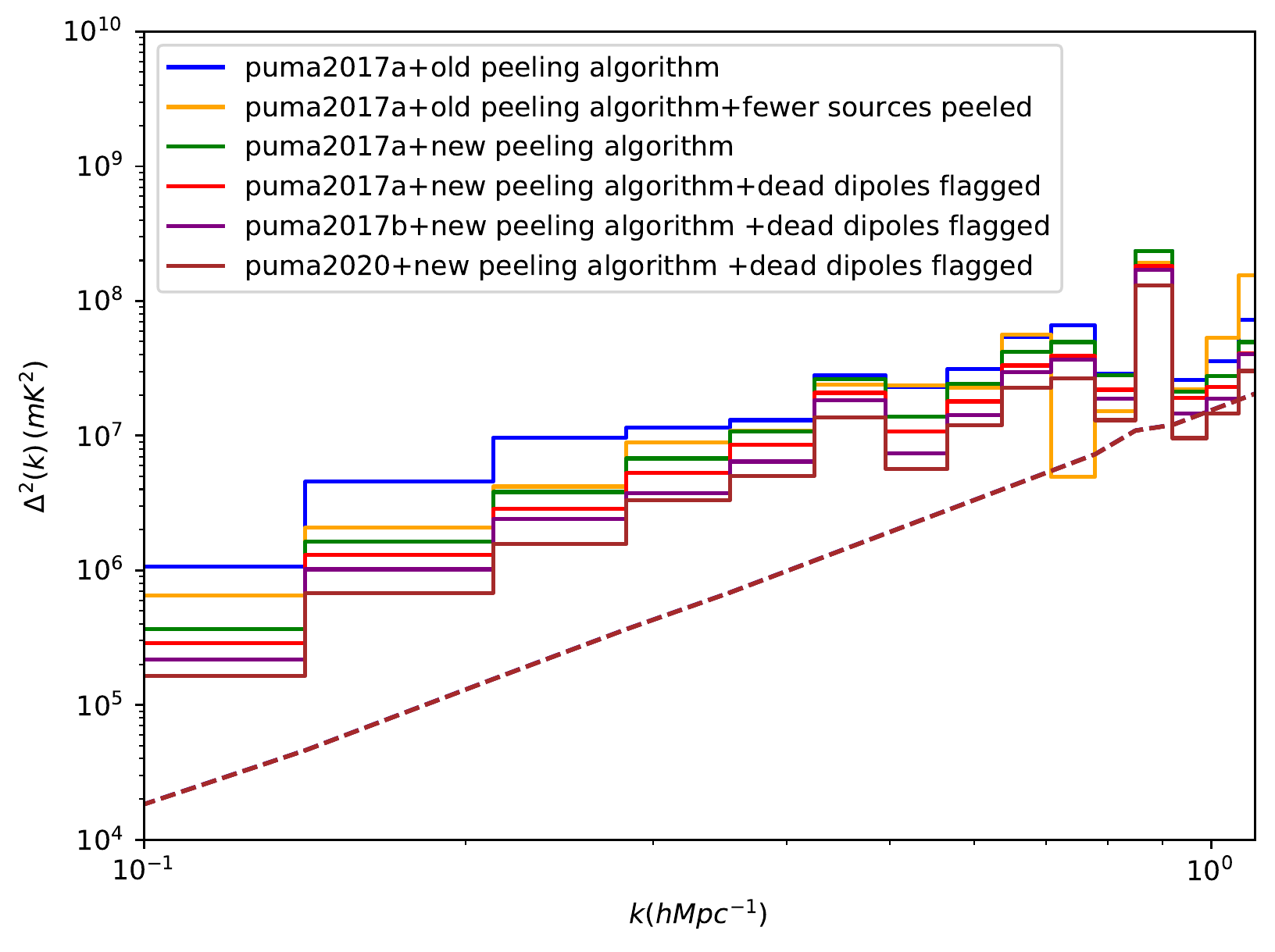}
    \caption{The improvement of data reduction as shown by comparing 1D power spectrum of one typical zenith observation processed by different versions of the source catalogue and calibration pipeline. The dashed line shows the thermal noise curve.}
    \label{fig:src-power-comp}
\end{figure}

We start off by showing the 1D PS for a single typical observation processed with the puma2017a source catalogue (as described in section \ref{sec:catalogue}). This processing is similar to that performed by \citet{Trott2020} and is shown as the blue line in Figure \ref{fig:src-power-comp}. 

The first improvement is due to the peeling algorithm. In order to effectively remove foreground sources, we set the RTS to peel the 1000 brightest sources in the field, in the order of their apparent brightness. However, due to a numerical issue that only affected peeling of faint sources, peeling fewer sources (orange power spectrum) would result in less contamination in comparison to complete 1000-source peeling (blue power spectrum). This difference is due to the fact that the peeling process with fewer sources does not go through the procedure of faint source peeling. Previous EoR1 processing in \citep{Trott2020} had some observations with fewer number of sources peeled.

Having resolved the numerical issue with the peeling of faint sources, the effective peeling of 1000 sources is reflected by a less contaminated power measurement shown by the green spectrum. Also, RTS processing typically treats the five brightest sources as full direction-dependant (DD) calibrators. Although reducing the number of DD calibrators has been shown to improve the results for Ultra-low data (Yoshiura et al., submitted), the results were inconclusive for High band data.

Another improvement is based on the availability of dead dipole information from 2014 onward. Dead dipoles occur when the low-noise amplifier on one of the cross dipoles fails. RTS forms the beam model for each tile individually. So the statistics of dead dipoles of each single tile provides us with the information to adjust the beam profile of the tile and the visibility weights. By updating the beam model for each tile, we take into account the effect of any dead dipoles, if it contains any. Without this information, additional sources of error are included \citep{Joseph2020}. Therefore, by flagging the dead dipoles within each tile, we obtain a less contaminated power spectrum as shown by red line.

Finally upgrading the sky models further improved the results by mitigating the foreground contamination. The purple line is the result of processing with puma2017b which includes updated sky model and the brown line is the result of processing with the puma2020 catalogue which includes the upgraded shapelet model of Fornax-A based on its Phase-I/II image (refer to section \ref{sec:catalogue}). Puma2020 catalogue is the basis of this work. There is $\sim6$ times improvement in power measurement at $P(k=0.1~\mathrm{h Mpc^{-1}})$  due to improvements in the  source catalogue and peeling process.

\section{Results}\label{sec:results}
In this section we demonstrate some of the prominent differences between the MWA EoR0 and EoR1 fields, as well between the 2015 data analysed here and the 2013 data analysed in some previous MWA results. We then present a series of data quality metrics used to identify contaminated data and select a best sample for power estimation. Our final power spectrum results are presented in section \ref{sec:limits}.

\subsection{Comparison of 2013 and 2015 Data}\label{sec:comp_13_15}
Previous MWA EoR data analysis works such as \citet{Dillon:2014}, \citet{Dillon:2015b}, \citet{Beardsley2016} and \citet{Barry2019} have been focused on data observed during 2013, while \citet{Li2019} and \citet{Trott2020} studied 2016 and 2013-2016 data, respectively.  In this work, we aim to process the data observed during 2015 (and 2014). Therefore, it is important to investigate the variability of data between the 2013 and 2015 seasons. We compare two datasets including nine zenith-pointing observations of a single night, observed in each of these two years. Both datasets are selected according to the metrics elaborated in section \ref{sec:metrics}.

\begin{figure}
    \centering
\subfloat[Visibility Noise RMS of 2013 (pink) and 2015 (aqua) data with their median values bold-ed in red and blue, respectively.]{
\includegraphics[width=0.9\columnwidth]{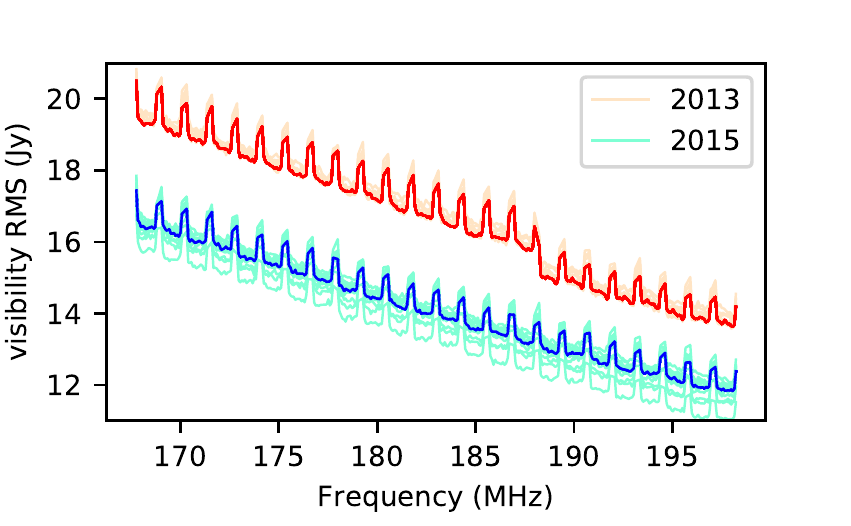}
  \label{fig:rms_2013_2015_a}
  }
\\
\subfloat[same as (a) while in case of 2015 the dead dipoles are not flagged.]{
  \includegraphics[width=0.9\columnwidth]{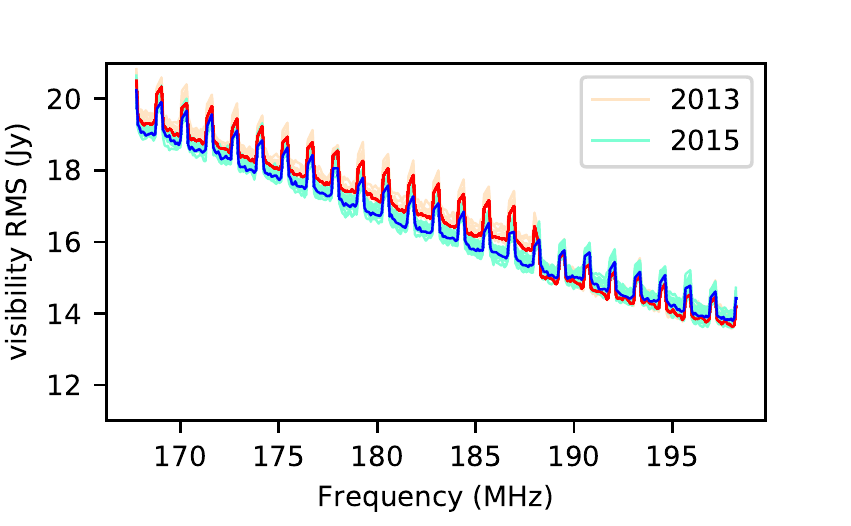}
  \label{fig:rms_2013_2015_b}
  }
\caption{Comparison of Visibility Noise RMS of 2015 and 2013 data.}
\label{}
\end{figure}
Figure \ref{fig:rms_2013_2015_a} illustrates the visibility noise RMS as a function of frequency of the two datasets, indicating that 2013 visibilities are more contaminated than 2015 ones. While the correlator temporal resolution of 2013 and 2015 data is $0.5~\mathrm{s}$ and $2~\mathrm{s}$ respectively, the main difference between the data of these two seasons is due to the recording of dead dipoles (as discussed in section \ref{sec:improvement}). There are $\sim$ 50 tiles with one dead dipole in 2015 data and the tiles with more than two dead dipoles are flagged \citep{Joseph2020}. For the 2015 observations, the beam shape and visibility weights are adjusted if there is a dead dipole recorded, but that is not the case with 2013 observations. Moreover, there is one local jump in visibility noise RMS values in 2013 observations which is due to a discontinuity in applied digital gains. This jump is not noticeable for the 2015 data where smoother digital gains are applied across the band. Also, increased RMS values are noticeable at the coarse band edges where digital non-linearities produce varying coarse band shapes.  
To investigate the effect of the application of dead dipole information, we disabled the feature of dead-dipole flagging in the 2015 data. In that case, as it is shown in figure \ref{fig:rms_2013_2015_b}, the RMS values of 2013 and 2015 observations generally agree with each other, though there are still some differences due to dissimilar dead dipoles in the two datasets.

\begin{figure}
    \centering
\subfloat[Ratio (left) and difference (right) of the power spectrum of 2015 versus 2013 data. In the difference plot, 2015 data has lower power than 2013 in red bins while it is higher in blue bins. The intensity of the difference between the two is indicated in ratio plot; As the difference gets larger, the bins tend to warmer colors of the scale.]{
  \includegraphics[width=0.9 \columnwidth]{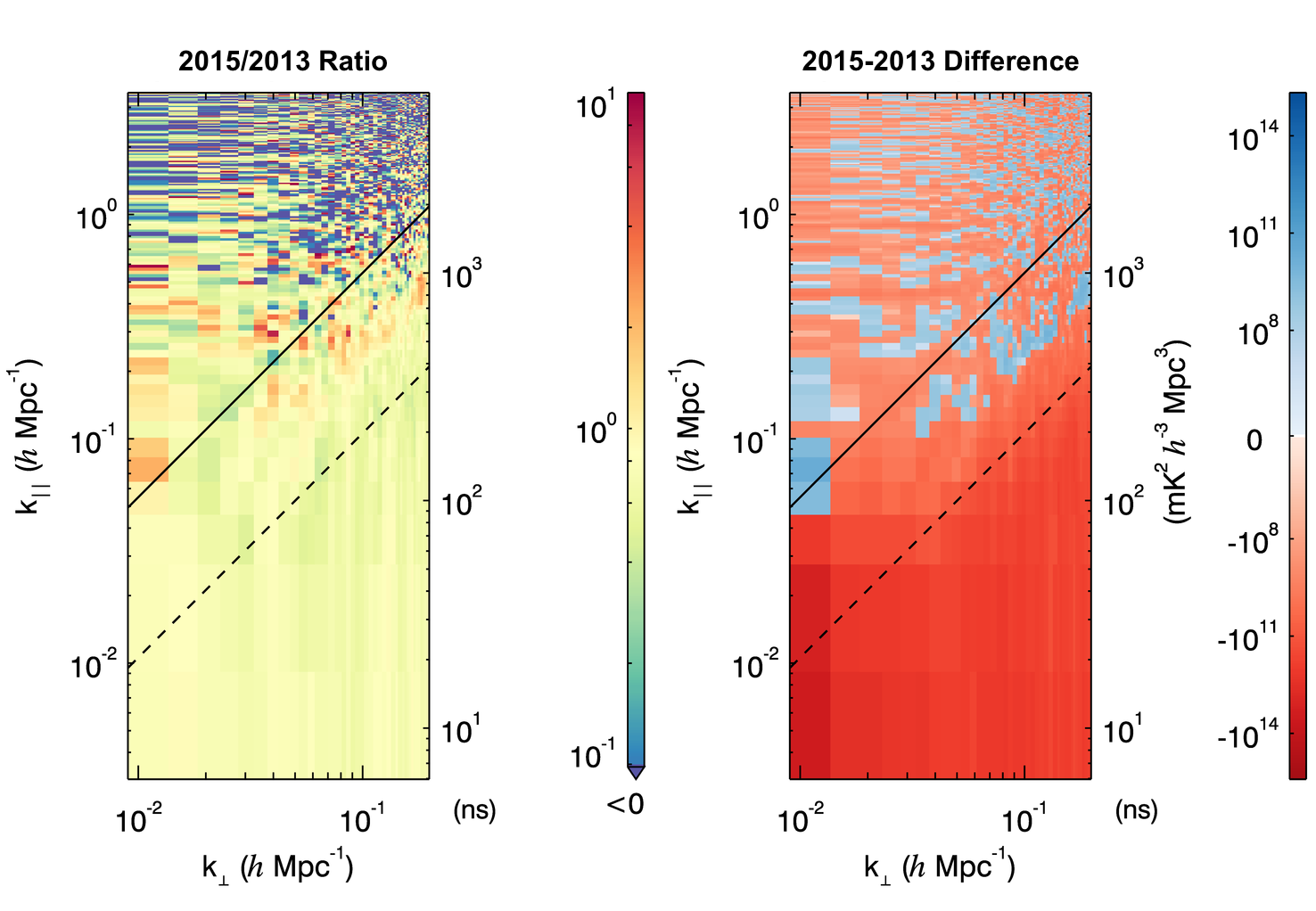}
  \label{fig:PS_13_15_2D}}
\\
\subfloat[1D power spectrum of 2013 (blue) data compared with that of 2015, both with dead dipole flagging enabled (orange) or disabled (green).]{
  \includegraphics[width=0.9\columnwidth]{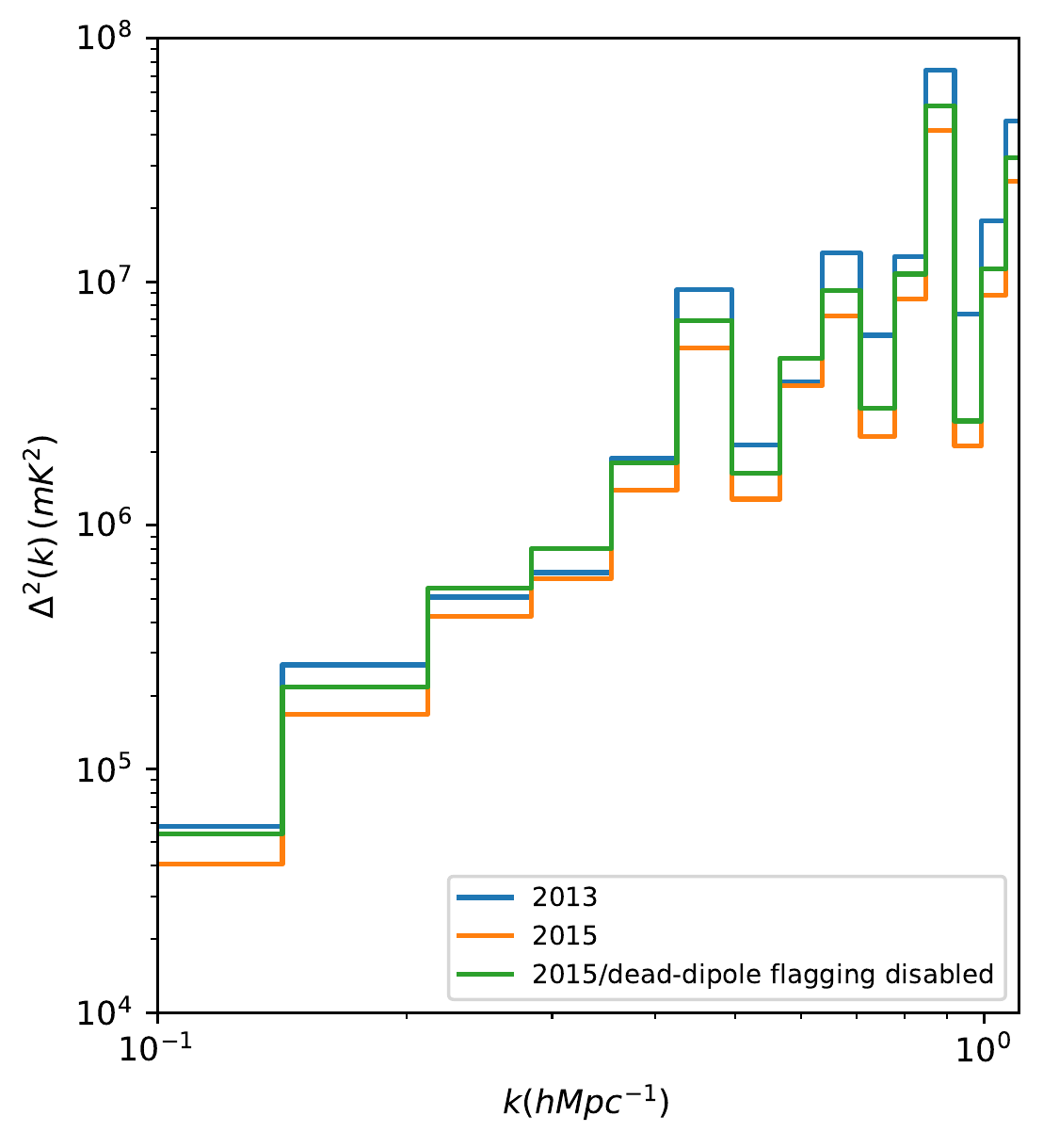}
  \label{fig:PS_13_15_1d}}
\caption{Power Spectrum Comparison between 2015 and 2013 data.}
\label{fig:PS_13_15}
\end{figure}

The effect of inclusion of dead-dipole information in 2D and 1D power space is shown in figure \ref{fig:PS_13_15}. The 2D power spectra comparison of 2013 and (dead-dipole flagged) 2015, shown in \ref{fig:PS_13_15_2D},  generally indicates less contaminated power measurements for 2015. Moreover according to figure \ref{fig:PS_13_15_1d}, while the 1D power spectrum of 2015 without dead-dipole flagging mostly agrees with that of 2013, it is noticeably (1.3 times at lowest k-modes) more contaminated than that of dead-dipole flagged data from 2015.

Both the visibility noise RMS and the power spectrum confirm there is less contamination in 2015 data relative to 2013 data, thereby reducing the systematic errors in this analysis of the 2015 data in comparison.

\subsection{Comparison of EoR0 and EoR1 Fields}\label{sec:comparison}
The MWA EoR0 field has been broadly analyzed and studied in previous works such as \citet{Dillon:2015b}, \citet{Beardsley2016}, \citet{Barry2019}, \citet{Li2019} and \citet{Trott2020}. However the EoR1 data has only been studied by \citet{Paul2016} and \citet{Trott2020}, with the former analyzing a 3-hour integration with the CASA/Delay-Spectrum pipeline and the latter analyzing $\sim$19 hours of zenith-pointing observations using the RTS/CHIPS pipeline. \citet{Zhang2020} also studied the performance of different calibration methods on MWA Phase-II data from EoR0 and EoR1 fields. In this work we aim to analyze EoR1 data with the RTS/CHIPS pipeline over a wider range of beam-pointings with an improved version of the RTS, higher fidelity source catalogues and improved data quality metrics. Before focusing on EoR1, we consider some of the main differences between these fields. The sky temperature of the two fields are also compared in appendix \ref{sec:sky-temp}.

\subsubsection{Per Pointing Comparison}
One of the underlying assumptions of all EoR experiments is that the calibrated visibilities are stable in time, allowing for long integrations which can attain the sensitivity required to detect the faint cosmological signal. However, as shown in \citet{Beardsley2016} and \citet{Barry2019}, there are prominent differences visible between the different MWA pointings for EoR0. In this section, we present differences visible between different pointings both for EoR0 and EoR1, and across different nights for EoR1. We illustrate these differences through the median of the visibility noise RMS (RMS hereafter) and the IonoQA (defined in section \ref{sec:data-metrics}).

\begin{figure*}
    \centering 
\subfloat[]{
  \includegraphics{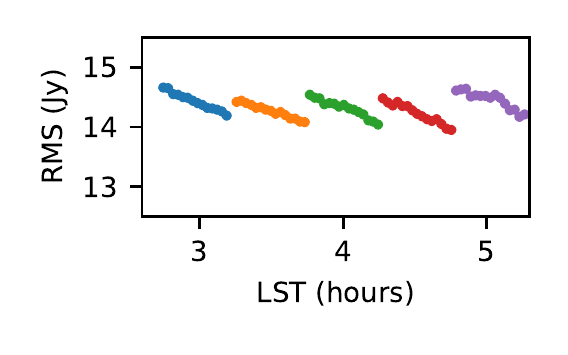}
  \label{fig:7}
}
\subfloat[]{
  \includegraphics{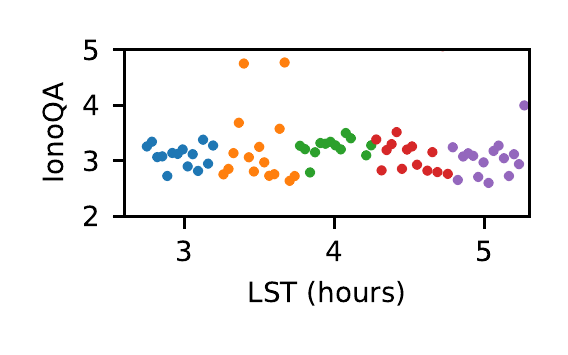}
  \label{fig:8}
}
\subfloat[]{
  \includegraphics[width=0.33\linewidth]{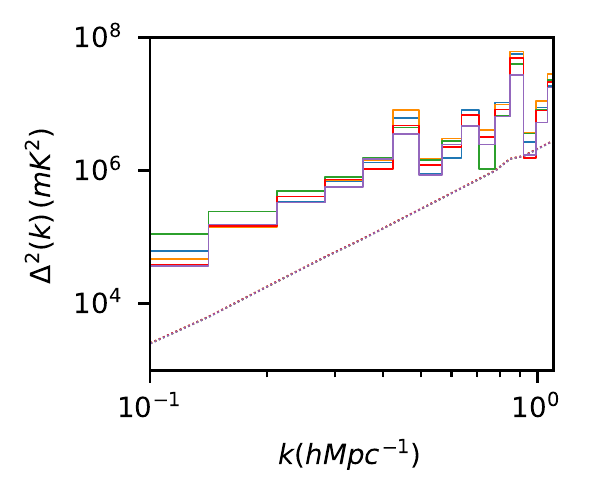}
  \label{fig:9}
}
\\
\subfloat[]{
  \includegraphics{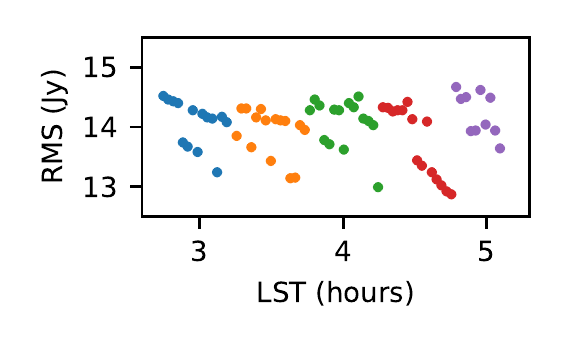}
  \label{fig:4}
  }
\subfloat[]{
  \includegraphics{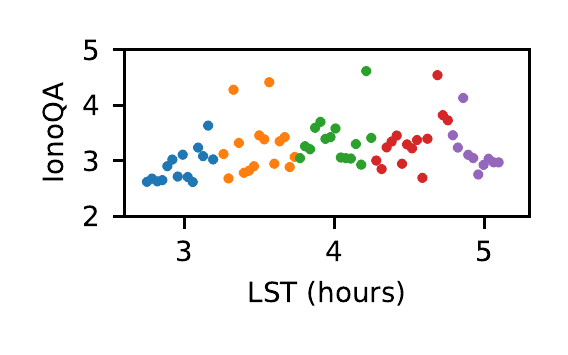}
  \label{fig:5}
}
\subfloat[]{
  \includegraphics[width=0.33\linewidth]{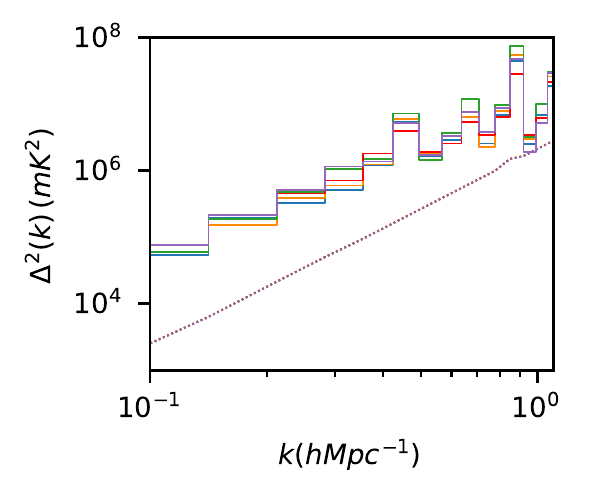}
  \label{fig:6}
}
\\
\subfloat[]{
  \includegraphics{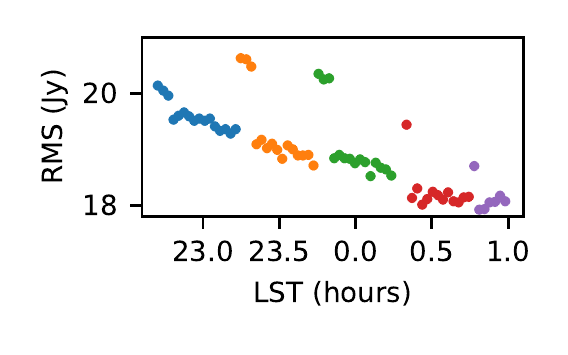}
  \label{fig:13}
}
\subfloat[]{
  \includegraphics{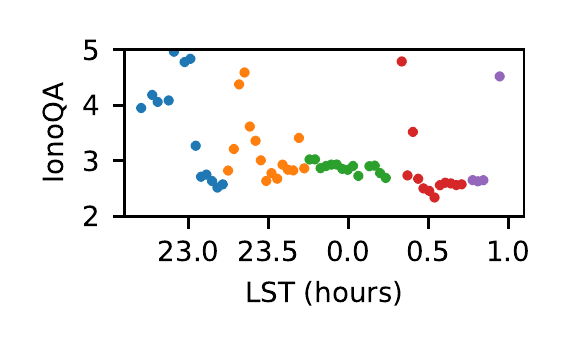}
  \label{fig:14}
}
\subfloat[]{
  \includegraphics[width=0.33\textwidth]{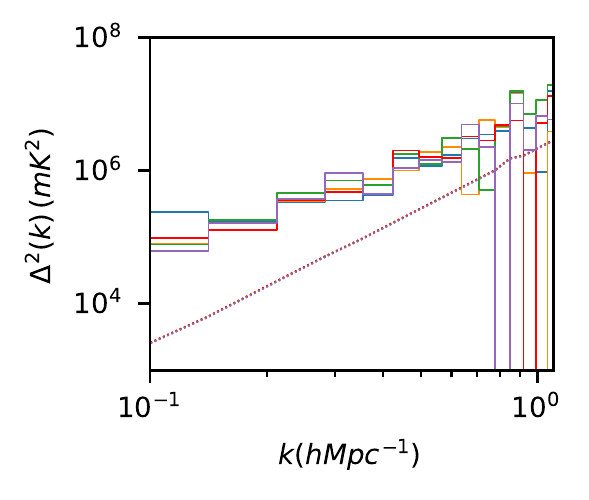}
  \label{fig:15}
}
\caption{The columns show the median RMS, IonoQA and 1D power spectrum, respectively. The first and second rows show the EoR1 observation nights  September 24 and 12 2015, respectively, while the third row shows the EoR0 observation night September 24 2015. Each data point indicates one observation and is color-coded by pointing: Minus2 (blue), Minus1 (orange), Zenith (green), Plus1 (red) and Plus2 (purple)}. The dashed lines in power spectra plots represent the thermal noise.
\label{fig:sep-nights}
\end{figure*}

Figure \ref{fig:sep-nights} displays the per-observation RMS and IonoQA, as well as the per-pointing 1D power spectra of three representative nights. Two nights (2015 September 24 and 12) from EoR1 are presented in the first two rows, and the third row presents a night (2015 September 24) from EoR0. 

Comparing the EoR1 nights, September 24 (first row) has a linear and predictable variation of RMS within each pointing transition. We interpret this as the primary beam repeatedly passing over regions of varying galactic emission. For the September 12 data (second row) the behaviour is more complicated; the linear, periodic structure still appears but some unknown effect is also introducing additional structure within each pointing. There appears to be no significant difference in the IonoQA values for these nights.

For the EoR0 observations from September 24 (third row), we observe an overall decline in the RMS values over the course of the night. We interpret this as lower sky temperature associated with the setting of the Galactic Centre, as predicted in Figure~\ref{fig:gsm}. The overall RMS for EoR0 observations is also higher than for EoR1, as discussed in 
section \ref{sec:sky-temp}. For this night, the IonoQA also appears to show some larger variations than seen in the EoR1 data later in the night. 

A small number of observations with IonoQA value of greater than 5 do not show up in the IonoQA plot as it is limited to quality assurance criteria \ref{sec:metrics}.

We note that the 1D power spectra of different pointings over these typical nights do not show a systematic intra-pointing variability either on different nights or on different spatial scales.

\begin{figure}
    \centering
\subfloat[]{
  \includegraphics[width=0.9\columnwidth]{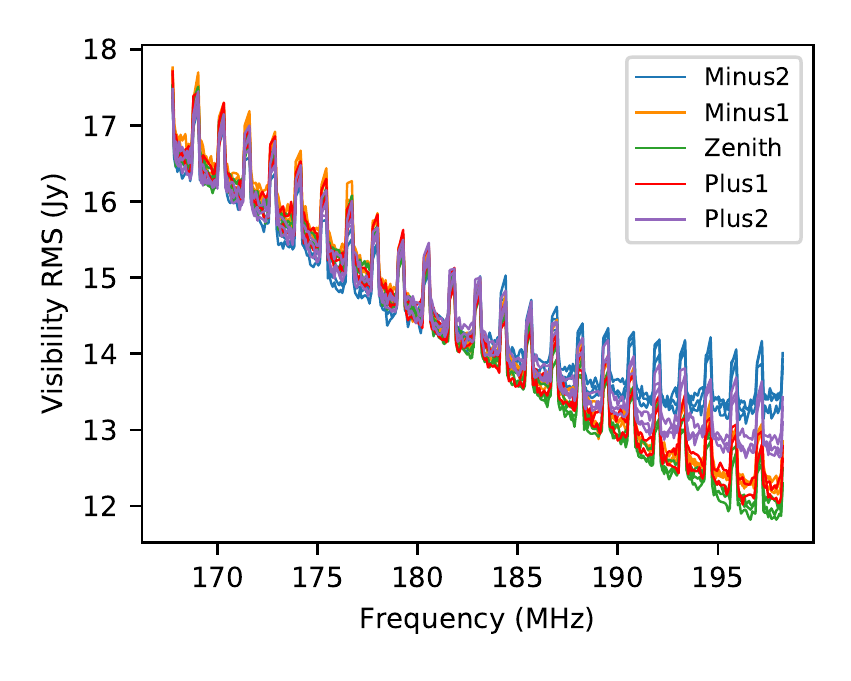}
  \label{fig:eor1-rms-10sep}
}
\\
\subfloat[]{
  \includegraphics[width=0.9\columnwidth]{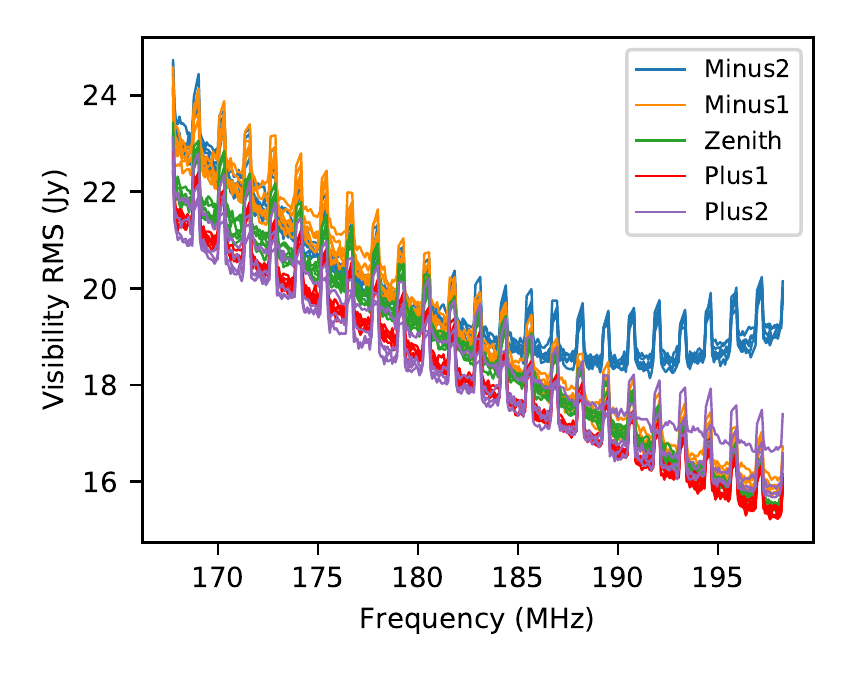}
  \label{fig:eor0-rms-10sep}
}
\caption{RMS over frequency per pointing of a) EoR1 and b) EoR0.}
\label{fig:rms-per-pointing}
\end{figure}

Figure \ref{fig:rms-per-pointing} shows the RMS across the whole frequency band for three observations from each pointing of EoR1 (top panel) and EoR0 (bottom panel). The RMS generally declines as we move towards higher frequencies as expected from the lower sky temperature. However, in the case of off-zenith pointings, the RMS does not follow a uniform trend and tends to slightly increase at high frequency. The slope increases for observations further from the zenith, even more so for the Minus3 and Minus4 pointings (not presented in this figure). As discussed in \citet{Beardsley2016}, although the MWA EoR fields are primarily devoid of Galactic plane contamination, the sidelobes of the beam are highly contaminated by Galactic emission when it rises above the horizon, one of the complications of wide-field instruments. This condition is significantly worse in before-zenith pointing transitions such as Minus4 to Minus1. EoR1 shows widefield contamination on either side of zenith, albeit at a lower level.

Per-pointing analysis of EoR1 data shows that the five central pointings are behaving consistently and thus approved for further analysis to measure the limits. 

\subsubsection{Integrated 2D Power Spectrum Comparison}
\begin{figure}
    \centering
\subfloat[2D power spectrum of EoR1 (left panel) and EoR0 (right panel) for 30 observations.]{
  \includegraphics[width=0.95\linewidth]{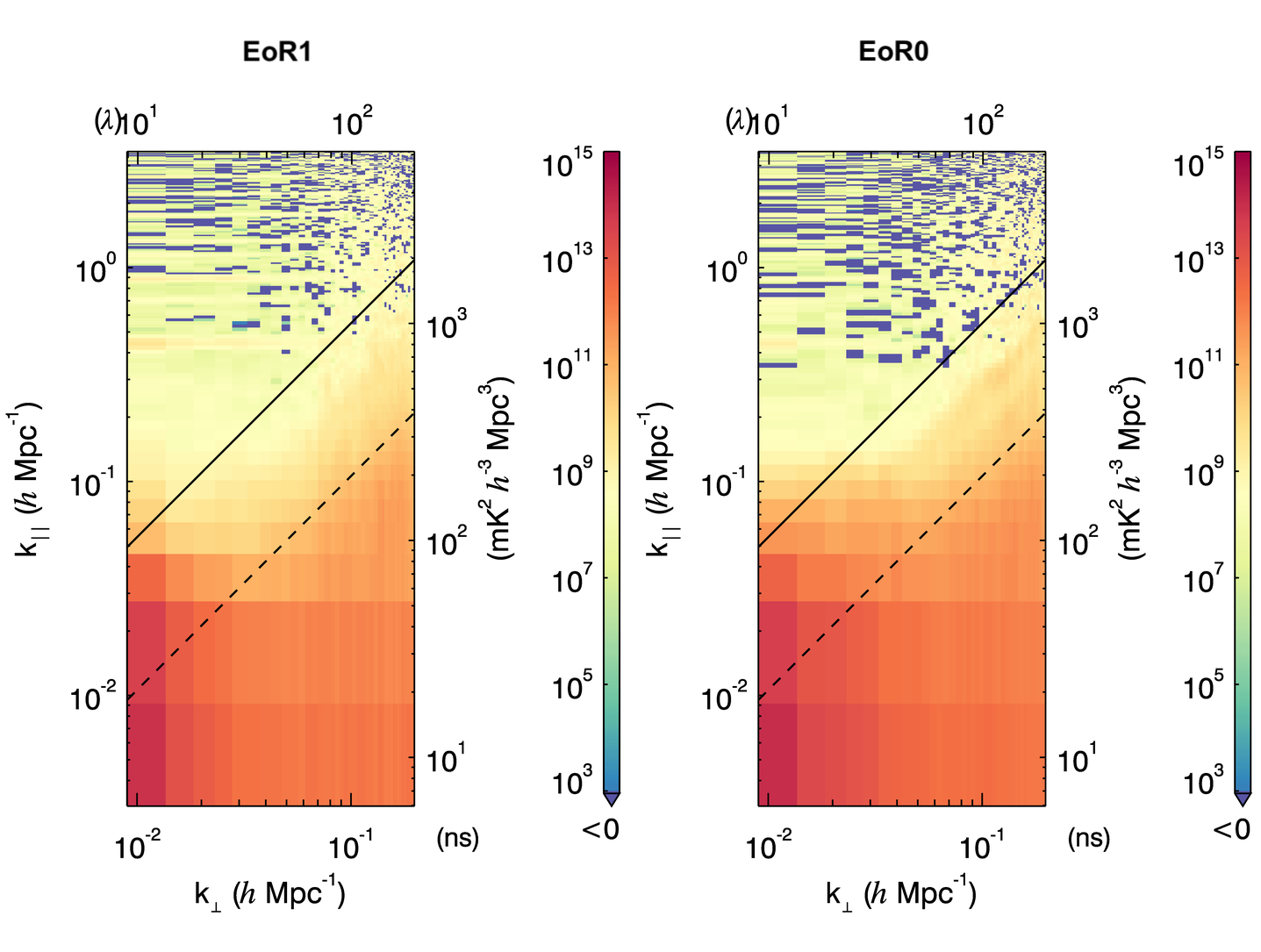}
  \label{fig:2d-raw-ps-eor1-eor0}
}
\\    
\subfloat[Ratio (left panel) and difference (right panel) of 2D power spectrum of EoR1 vs. EoR0.]{
  \includegraphics[width=0.95\linewidth]{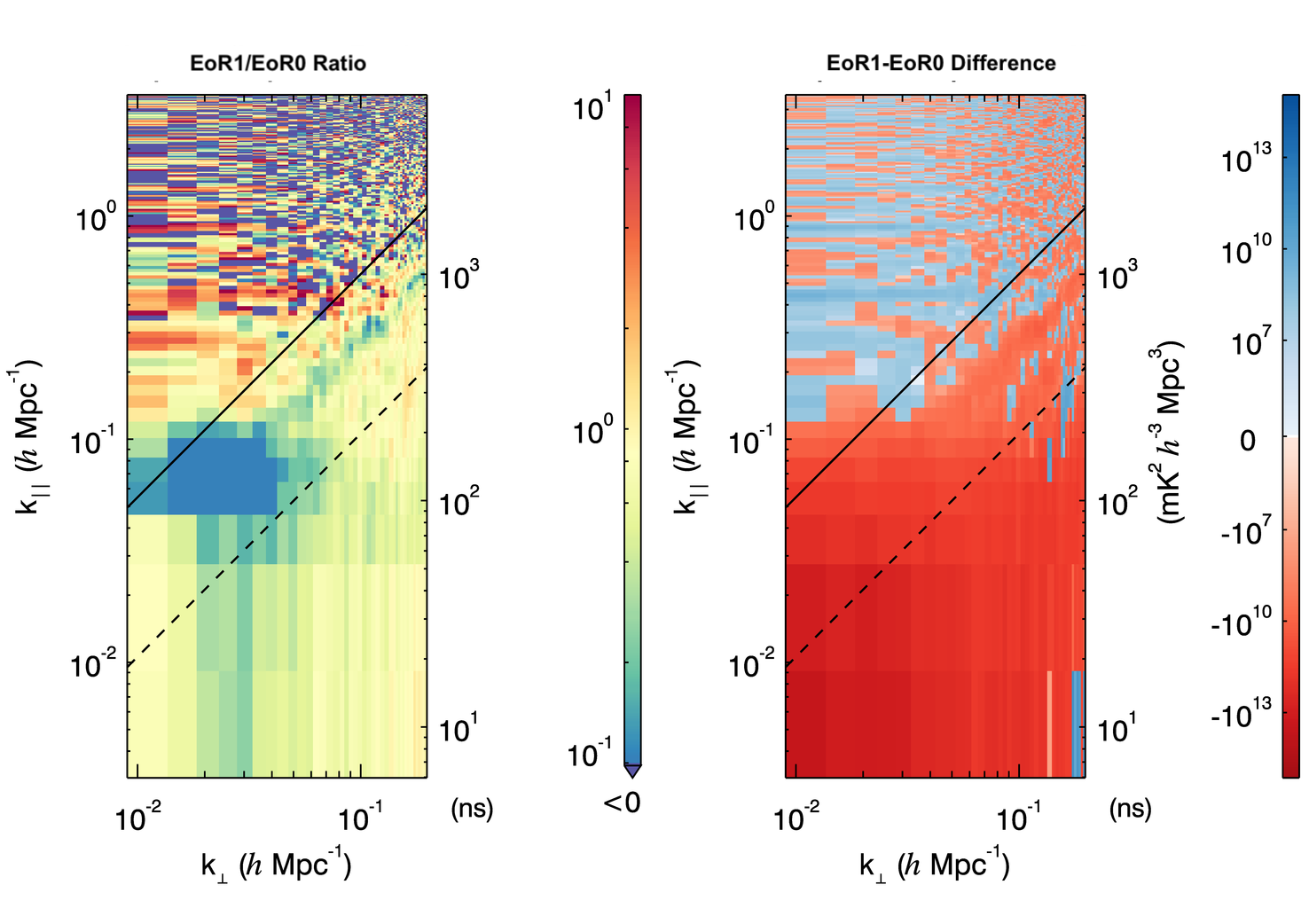}
  \label{fig:2d-ps-eor1-eor0}
}
\\
\subfloat[Measured 1D power spectrum comparison of EoR1 (solid blue) and EoR0 (dashed red). Thermal noise is plotted as dotted blue/red line for EoR1/EoR0 fields.]{
  \includegraphics[width=0.95\columnwidth]{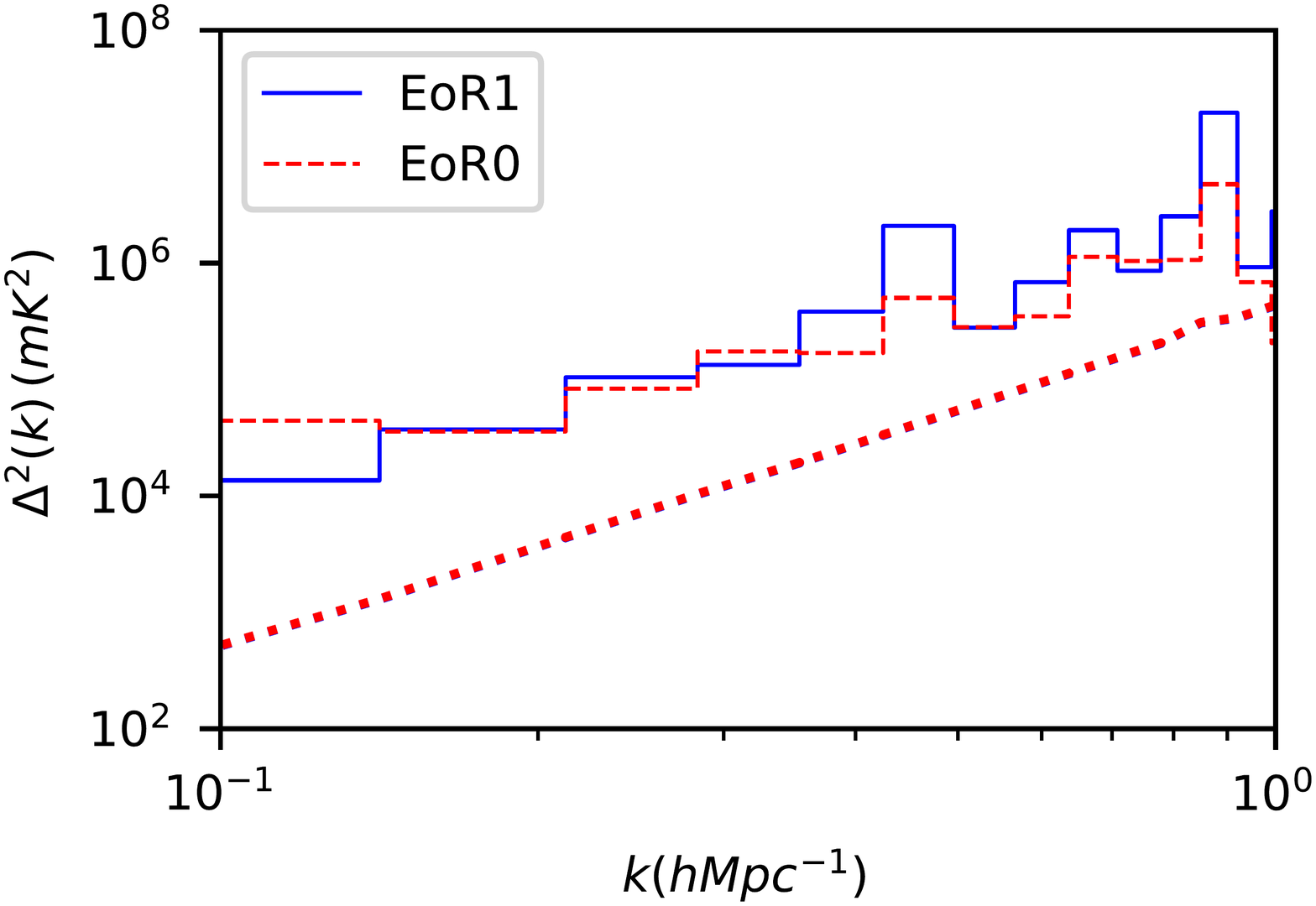}
  \label{fig:2d-ps-eor1-eor0-jan21}
}
\caption{EoR1 and EoR0 power spectrum (top panel) and their comparison in 2D (middle panel) and 1D (bottom panel) power space.}
\label{fig:eor0-eor1-ps}
\end{figure}

The power spectra of a $\sim$ 1-hour integration of EoR1 and EoR0 (6 observations from each of five central pointings) are compared in figure \ref{fig:eor0-eor1-ps}. The 2D power spectrum comparison (top panel) indicates higher contamination in EoR window and so generally higher 1D power (bottom panel) for EoR1. However, the power in the wedge is higher for EoR0 and is particularly noticeable along the horizon in the ratio plot (middle panel).
\citet{Thyagarajan:2015} demonstrated that the wide-field measurements are highly sensitive to diffuse emission at the horizon. This feature is due to the shortening of baselines near the horizon which makes them more sensitive to larger angular scales. They also subtend larger solid angles near the horizon and so capture a larger integrated emission. Therefore, we expect Galactic emission, which is mostly present at off-zenith transitions of EoR0 \citep{Trott2020}, to introduce higher contamination if the foreground modelling and removal is not perfect. This excess power is leaking into large spatial scales or equivalently small $k_\perp$ areas of the horizon in the ratio plot, and is captured by lowest bin of 1D power spectrum. This difference shows up as lower EoR1 power at lowest spatial modes, or equivalently large scale measurements in 1D power spectrum (bottom panel).

Having improved the analysis tools, that would be promising to perform further deep measurements of EoR1 field and validate its capability to less contaminated limits in comparison to EoR0, at least at large scales. 
\subsection{Data Quality Metrics}\label{sec:metrics}
All 767 candidate observations were successfully calibrated. In order to select the best dataset for power spectrum analysis, we need to filter out any highly contaminated observations. We combine a number of criteria applied by previous MWA EoR analyses together with novel methods of identifying anomalous data. Specifically, our selection is performed over six stages, as follows: (i) auto-correlation amplitudes, (ii) calibration-based metrics, (iii) window/wedge power metric, (iv) individual 1D power spectrum, (v) ultra-fine RFI, (vi) Phase Component of Gain Profile. Each one is described below in more detail and summarised in table \ref{tab:metric-stat}.
\begin{enumerate}
\item Auto-Correlation Profiles

Examining the (time-averaged) auto-correlation amplitudes of individual tiles readily identifies features such as faulty unflagged tiles or digital TV (DTV) interference. Since the MWA is fairly stable with well-behaved auto-correlations, inferences about extreme outliers and large changes in frequency can be made even prior to calibration. The auto-correlation amplitudes generally vary smoothly over frequency. For all tiles with the same cable-group\footnote{It refers to beamformer-to-receiver cable group.}, the amplitudes are distributed with a maximum deviation of roughly $20\%$ around the mean value. However some tiles are extreme outliers or have large deviations over short frequency-scales as high as $25-30\%$ of the mean value. Thus the associated observations were removed from the dataset.

 To learn more about the large variations appearing in some tiles, we measured the standard deviation of the auto-correlation amplitudes over time as a function of frequency. We noticed that some tiles, specifically 55, 116, 125 and 163, show large deviations in some frequency channels. The detected standard deviation values might be as large as those produced by DTV, as illustrated in figure \ref{fig:auto}, and are caused by a sharp peak at a specific time/frequency of the amplitude profile. Since such features are not flagged by AOFlagger, we probed their effect on power measurements by flagging these tiles. Their effect on the measured power spectra is not noticeable at this stage; implying that the level of their introduced systematics is below the sensitivity of current analysis. However, with future development of higher precision techniques and larger datasets, removing these observations from the data might reduce the systematic floor.
    \begin{figure}
     \centering
     \includegraphics[width=0.99\columnwidth]{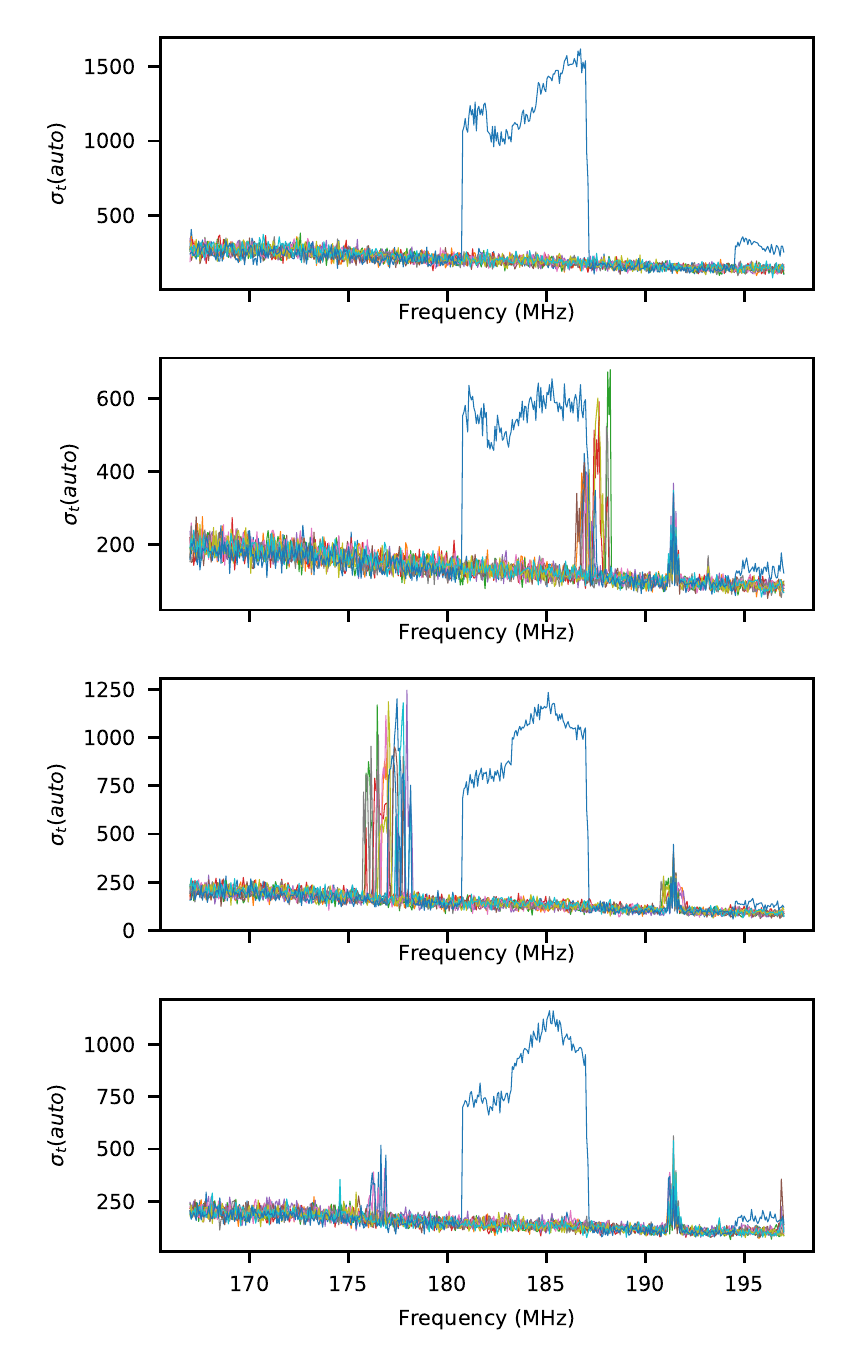}
     \caption{The standard deviation of the auto-correlation amplitude over time as a function of frequency for tiles 152, 163, 125 and 55 (respectively from top to bottom), colour-coded for 11 observations. Tile 152 is an example of a healthy tile, while the other three tiles show spikes at some frequency channels. The observation plotted in blue is DTV interference.}
     \label{fig:auto}
    \end{figure}

\item Calibration-Based Metrics

At this stage, all of observations are probed in terms of:
\begin{itemize}
    \item Calibration Solutions
    
    The RTS-derived bandpass and gain solutions for each observation are examined. In general, the amplitude and phase components of these complex quantities are consistent for all the tiles within one observation. Observations with at least one tile with an abnormal profile, typically large or sudden variations in their gain amplitudes are excluded. In addition, observations with more than 6 flagged tiles are removed.
    
   \item Ionospheric Activity
   
    Based on the studies by \citet{Jordan2017, Trott2018}, the ionospherically active observations with large offset and anisotropy measure should be avoided. So we exclude observations with IonoQA greater than 5.
    
    \item Visibility Noise RMS
    
    Visibility RMS (discussed in section \ref{sec:data-metrics}) generally varies smoothly over frequency. However,  highly contaminated observations whose profile fluctuation over at least one fine channel was larger than $40\%$ of the overall RMS change over the whole band were excluded. 
\end{itemize}

\item Window/Wedge Power Metric

Next, we apply the window/wedge metric as developed by \citet{Beardsley2016} and employed in previous studies such as \citet{Trott2020}. This is measured by a delay transform power spectrum estimator, computed incoherently across all short baselines and treats each baseline separately. Window power estimates the power in a selected region of the EoR window; between the main beam lobe and the first coarse channel harmonic. Wedge power measures the power in the primary beam main lobe wedge. The cuts on window power pick values within $2\sigma$ range of the distribution, as shown in the figure \ref{fig:wp_a}. The cut is also shown in the scatter plot of window power versus RMS in figure \ref{fig:wp_b}. We also apply $2\sigma$ cut on wedge power as shown in figure \ref{fig:window_wedge}. 
\begin{figure}
    \centering
    \subfloat[Histogram of window power and $2\sigma$ selection range of window power values marked by dashed lines.]{
        \includegraphics[width=0.9\columnwidth]{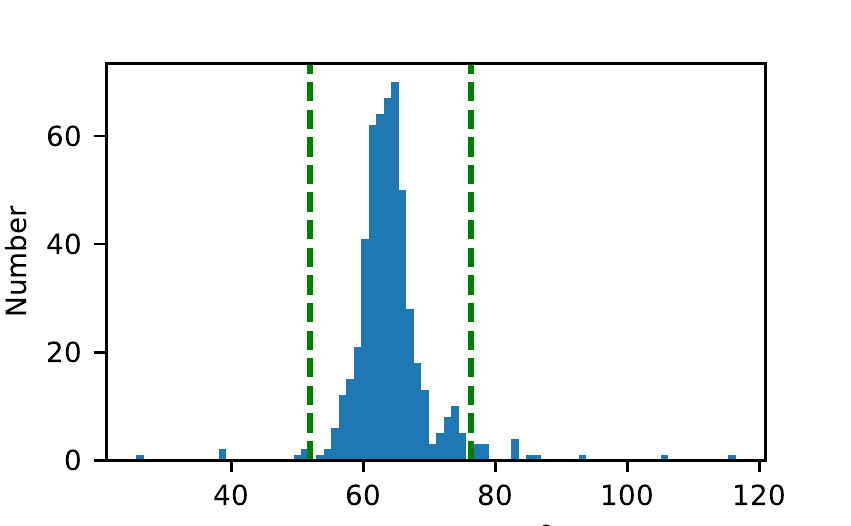}
       
        \label{fig:wp_a}
    }\\
   \subfloat[Distribution of window power versus RMS and selected range of values marked by dashed lines.]{
        \includegraphics[width=0.9\columnwidth]{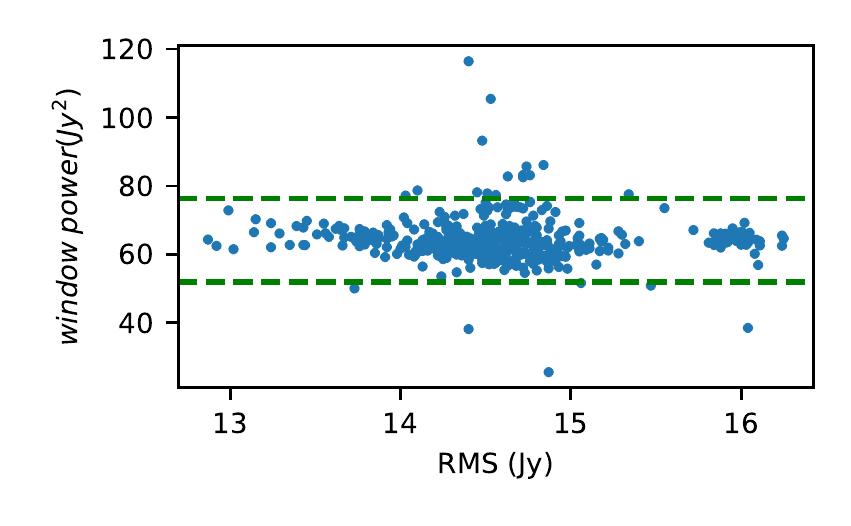}
       
        \label{fig:wp_b}
   }\\
   \subfloat[Distribution of window power versus wedge power and selected range of values marked by dashed lines.]{
        \includegraphics[width=0.9\columnwidth]{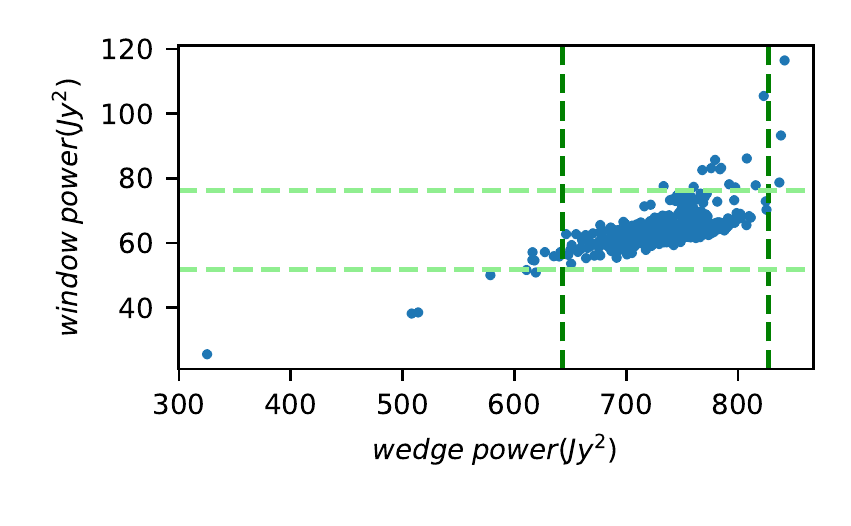}
        \label{fig:window_wedge}
   }
    \caption{Window/Wedge power cut-off}
    \label{fig:wp}
\end{figure}

\item Individual 1D Power Spectra

Having examined the incoherent delay spectrum of observations in the last step, we need to verify their power spectra which is a coherent measurement of the gridded visibilities. Hence, the 1D power spectrum of each observation is individually measured, for consistency with previous work \citep{Trott2020}. Figure \ref{fig:hist_p_k1} shows the histogram of $P$(lowest-bin) for candidate observations. The long tail of this histogram shows the clearly contaminated observations with high power measurements raised by some systematics. We apply Chauvenet's Criterion to detect the outliers. This method cuts the outliers with $P(k=0.1) > 1.35\times 10^5 \mathrm{mK}^2$ in the long tail, as shown with dashed red line in the figure. We examined the effect of their inclusion in the whole dataset power spectrum and see a drop in the total power implying they are clearly contaminated and need to be excluded from the total set. 
\begin{figure}
    \centering
    \includegraphics[width=\columnwidth]{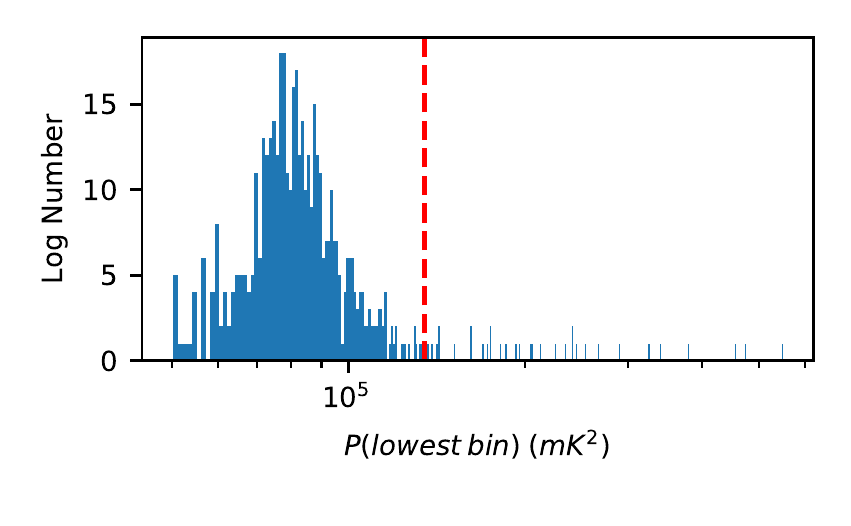}
    \caption{Histogram of P(lowest bin) for candidate observations.}
    \label{fig:hist_p_k1}
\end{figure}

Having cleaned the dataset by applying the last four metrics, 459 observations are selected. We can be even more conservative by considering two more subtle metrics, as follows:

\item Ultra-Faint RFI

    As described in section \ref{sec:preprocess}, standard MWA RFI flagging is handled by AOFlagger in this work. However, we are able to further detect significant ultra-faint RFI contamination occurrences in the data using SSINS. While we expect a Gaussian distribution for the mean-subtracted noise spectrum of a non-contaminated observation, as illustrated in figure \ref{fig:ssins-healty}, the most repeated occurrences of contaminated cases are either bright DTV interference as in figure \ref{fig:ssins-dtv} or a time-variant source of interference present over the whole observation time lapse and frequency band as shown in figure \ref{fig:ssins-twotier}. In total, 120 observations with abnormal SSINS profiles are identified by visual inspection. The abnormal cases include those illustrated in figures \ref{fig:ssins-dtv}, \ref{fig:ssins-twotier} and presence of narrowband RFI over time or frequency.

\begin{figure}
    \centering
    \subfloat[]{
        \includegraphics[width=0.9\columnwidth]{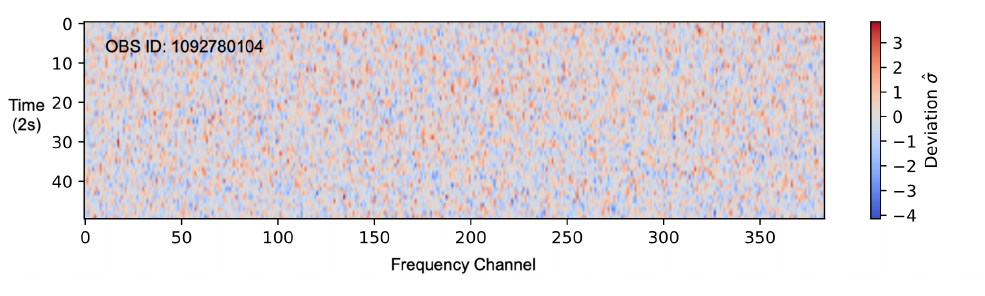}  \label{fig:ssins-healty}  
        }\\
    \subfloat[]{
        \includegraphics[width=0.9\columnwidth]{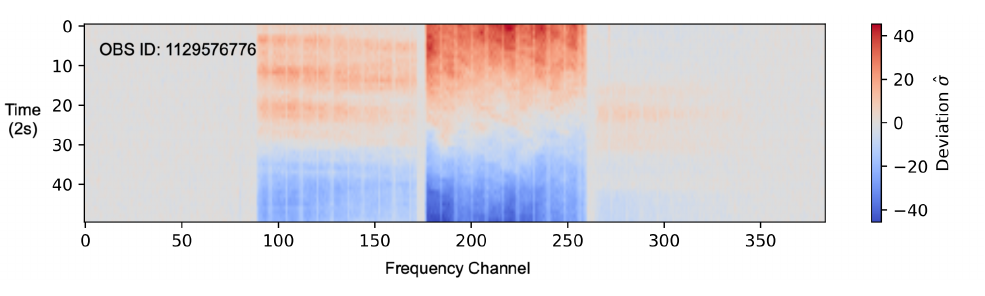}
    \label{fig:ssins-dtv}  
    }\\
    \subfloat[]{
        \includegraphics[width=0.9\columnwidth]{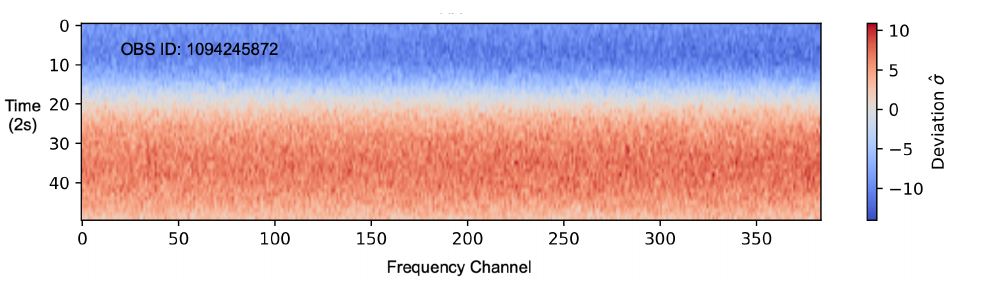}
        \label{fig:ssins-twotier}  
    }
\caption{Sky-subtracted Noise Spectrum of three observations with normal (top), DTV-interfered (middle) and biased (bottom) noise spectrum. Western Australia channel 6 ($177.5~\mathrm{MHz}$), channel 7 ($184.5~\mathrm{MHz}$) and channel 8 ($191.6~\mathrm{MHz}$) fall within MWA high-band frequency range.}
\label{fig:ssins}
\end{figure}

\item Phase Component of Gain Profile
    
    There are some observations which show excessive phase wrapping on the tiles associated with one or more receiver. This issue is detected in 118 observations.  
\end{enumerate} 

Having applied all six metrics, 221 observations are selected.

As the inclusion of the data detected by metric (v) and (vi) in the dataset does not make any noticeable change in the power spectrum, we could not verify if they are definitely a source of systematic contamination in the measurement at this stage. However they provide a cautiously refined dataset for further precise analysis of systematic contamination as discussed in next section \ref{sec:rms-based}.

\begin{table}
    \centering
    \begin{tabular}{clc}
    \hline
       metric  &  description & number of passed observation \\
       \hline
        * & successful calibration & 767 \\
        i & auto-correlation & 758 \\
        ii & calibration-based metrics: & 522 \\
         & solutions, IonoQA, RMS & \\
         iii & window/wedge power & 494 \\
         iv & individual power spectrum & 459 \\
         v & SSINS & 339 \\
         vii & phase component of gain & 221 \\
         \hline
    \end{tabular}
    \caption{Data quality metrics that were applied to the dataset in order and the associated number of passed observations. The final power spectrum is obtained from 459 selected observations that are selected by first four metrics. The last two metrics of v and vi are applied for further systematic analysis purpose.}
    \label{tab:metric-stat}
\end{table}
\subsection{Data Selection Based on Visibility RMS}\label{sec:rms-based}
\begin{figure}
    \centering
    \subfloat[1D power spectrum of a low-RMS dataset and a high-RMS one, each containing 54 observations.]{\includegraphics[width=0.9\columnwidth]{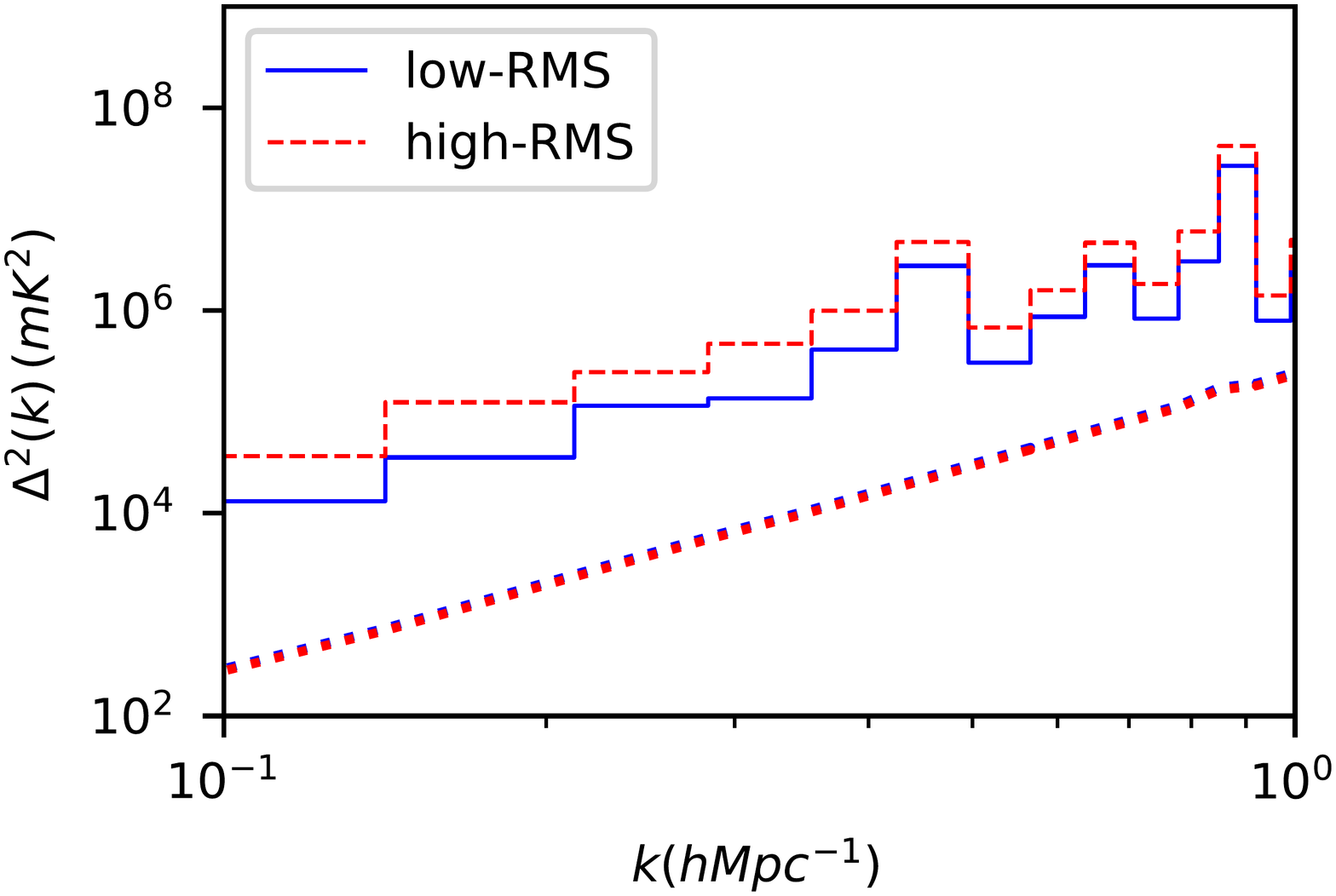}
    \label{fig:rms-low-high}}\\
    \subfloat[1D power spectrum of a low-RMS dataset and a mixed low/high RMS dataset, each containing 108 observations.]{\includegraphics[width=0.9\columnwidth]{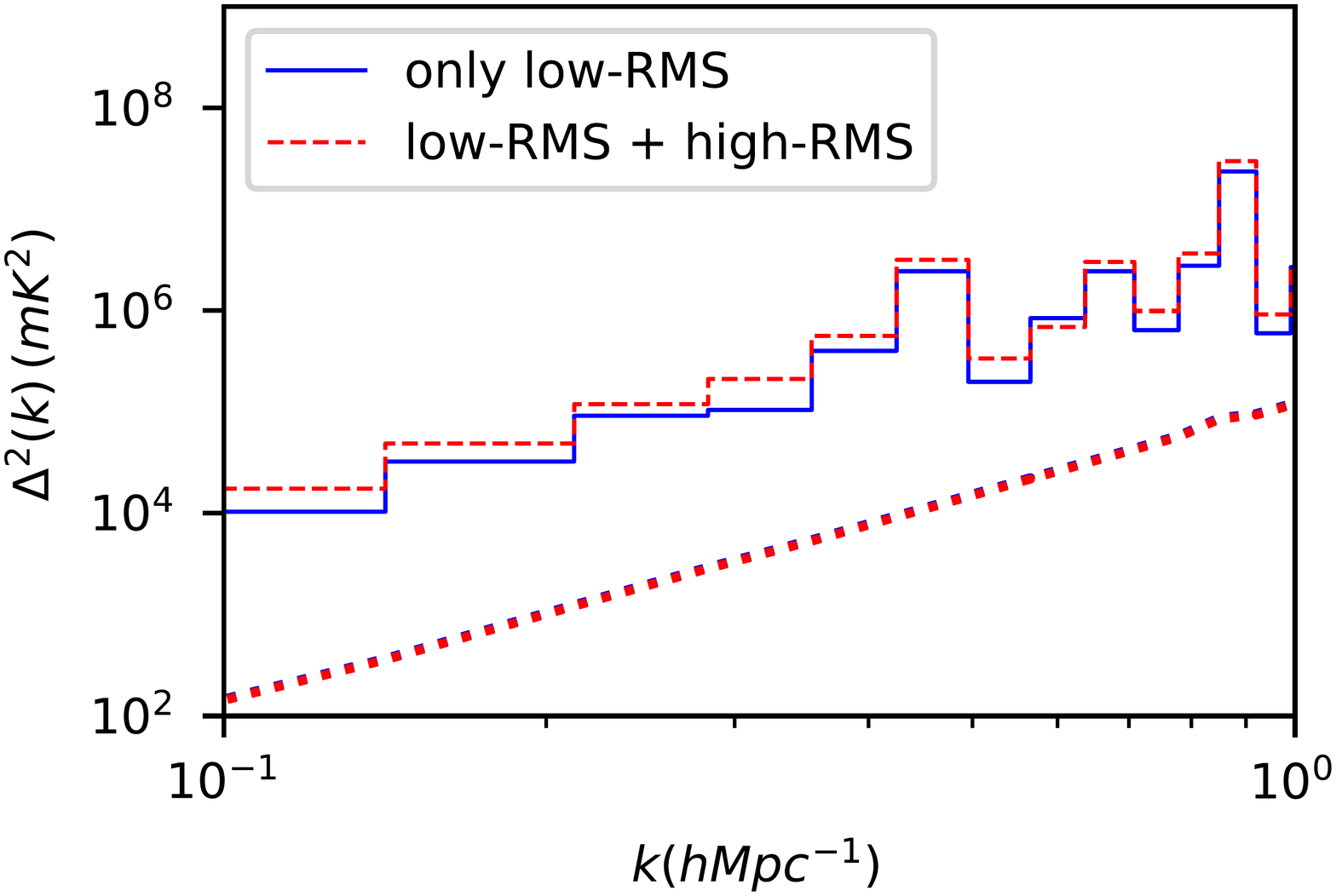}
    \label{fig:rms-low-mixed}}\\
    \subfloat[2D power spectrum of a low-RMS dataset versus that of mixed-RMS one.]{\includegraphics[width=0.9\columnwidth]{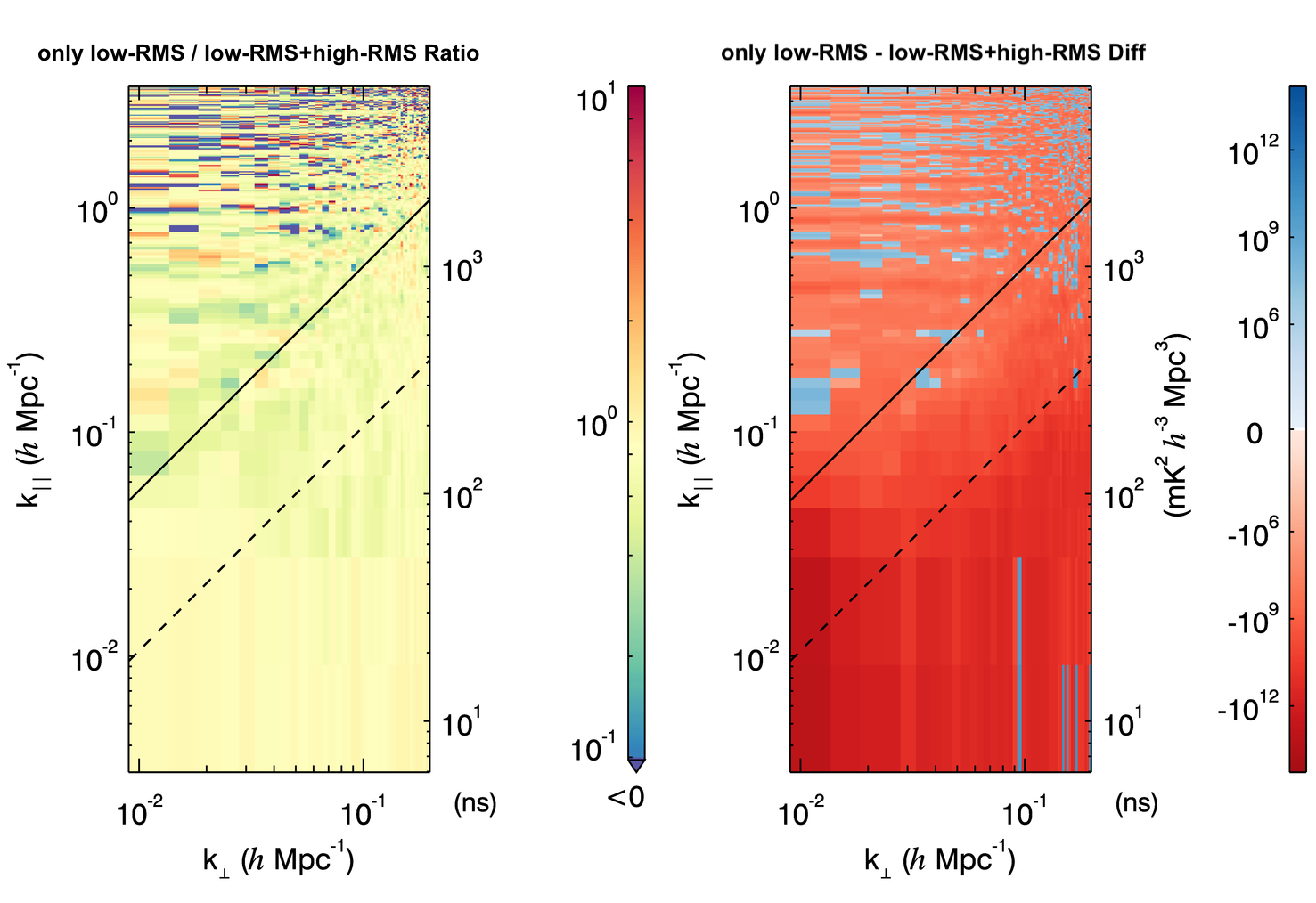}
    \label{fig:2d-low-mixed}}
    \caption{Power Spectrum comparison between low and high RMS observations}
    \label{figure:rms-compare-power}
\end{figure}

The main limitation on detecting the EoR signal at this time are the (un)known systematics. Visibility noise RMS could represent a combination of different systematics, as seen in the different receiver temperatures inferred in section~\ref{sec:sky-temp}. So we explore the residual systematic errors based on this metric. To make sure that our dataset is homogeneous, we perform this test only for the 221 observations that have been filtered by all the aforementioned six metrics. As shown in figure \ref{fig:wp_b}, for the 459 observations used in our final results, the median RMS values are centred at $\sim 14.8~\mathrm{Jy}$ within a range of $\sim 3~ \mathrm{Jy}$. By splitting the data into two groups of low-RMS and high-RMS with median RMS value of lower and higher than $14.8~\mathrm{Jy}$ respectively, we see significantly different results for their power measurements (figure \ref{figure:rms-compare-power}). Two datasets with the same number of observations from each RMS group were selected. The 1D power spectrum of low-RMS dataset is significantly lower than the high-RMS one, shown in figure \ref{fig:rms-low-high}. To investigate the effect of presence of high-rms data in the whole dataset, we also compared the power spectrum of a dataset including equal amounts of low-RMS and high-RMS data with another dataset containing only low-RMS data, shown in figure \ref{fig:rms-low-mixed} and \ref{fig:2d-low-mixed}. This demonstrates the increased level of contamination in the case of high-RMS visibilities. 

There are 167 low-RMS observations in the best 221-observation dataset. The lowest upper limits obtained from the low-RMS data is $\num{7.34e3}~\mathrm{mK^2}$ at $k=0.13~\mathrm{h~Mpc^{-1}}$ and z=6.5. Among the 221-observation clean dataset, the high-RMS observations mostly belong to a particular part of the observing season. For example, $100\%$ of 2015 September observations exhibit of low-RMS while all 2014 September and August observations are all of high RMS. As the high-RMS observations are concentrated in well-defined time-frames, they might be affected by some hardware artefact that occurred and lasted during that time. Having learned about these effects, a future approach might be to diagnose and avoid the observations of those faulty nights prior to mass processing.

\subsection{Final Power Spectrum Measurements}\label{sec:limits}
To avoid the cosmological evolution of the signal over the bandwidth of interest, we need to measure it within an interval which assures the signal ergodicity. We pick sub-bands of width $15.36~\mathrm{MHz}$ which will be tapered to an effective width of $\sim10~\mathrm{MHz}$ once a Blackman-Harris window function is applied. Hence, we will have three sub-bands with central redshifts of 6.5, 6.8 and 7.1, each including 192 fine frequency channels.

In the search for data  which is only noise-dominated, we explored six metrics to exclude highly-contaminated data. While the first four metrics are required to assure the data is not contaminated, we also conservatively applied the last two metrics to eliminate the possible finer signatures of systematics in the data. 459 observations passed the first four metrics but only 221 pass also the metrics (v) and (vi). Due to the relatively low integration time of selected or finely-selected datasets, the possible effects of detected systematics would not be evident in the final power measurement. However, the lowest  $2\sigma$ upper limits for 221-observation dataset is $\num{7.29e3}~\mathrm{mK^2}$ at $k=0.13~\mathrm{h~Mpc^{-1}}$ and z=6.5.

\begin{figure}
    \centering
    \includegraphics[width=0.9\columnwidth]{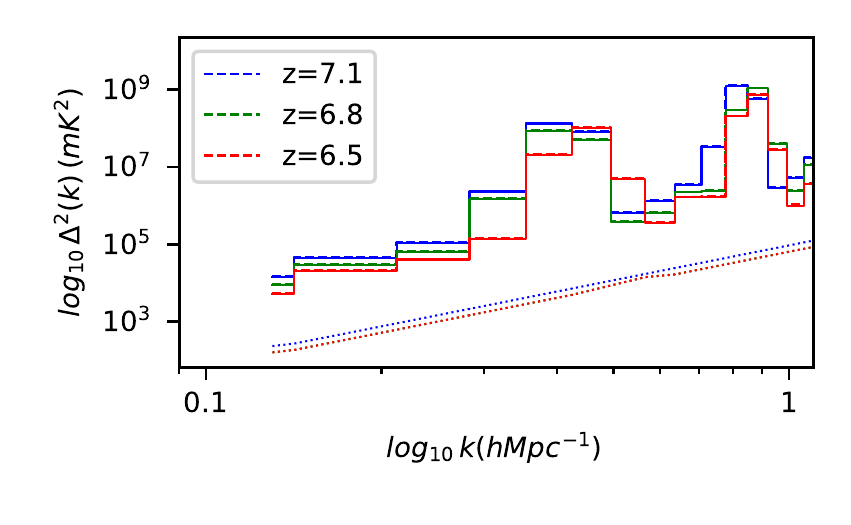}
    \caption{Measured 1D power spectrum (solid lines) and $2\sigma$ upper limits (dashed) for the 459 observation at z=6.5(red), z=6.8(green) and z=7.1(blue).}
    \label{fig:limit_459}
\end{figure}

We confirmed the accuracy of the power normalisation through a set of simulations. As is shown in appendix \ref{app:sim}, the simulated EoR signal is consistently recovered through RTS/CHIPS pipeline. The measured power and $2\sigma$ upper limits of 459 selected observations are presented in table \ref{table:limits}, and figure \ref{fig:limit_459}. This provides us with the lowest limit of $\Delta^2 \leq \num{5.44e3}~\mathrm{mK^2}$ at $k=0.13~\mathrm{h~Mpc^{-1}}$ and z=6.5.

\begin{table*}
  \centering
    \begin{tabular}{llllllllll}
    \toprule
          & \multicolumn{3}{|c|}{\textbf{z=6.5}} & \multicolumn{3}{c|}{\textbf{z=6.8}} & \multicolumn{3}{c|}{\textbf{z=7.1}} \\
    \midrule
    \textbf{k} & $\boldsymbol{\Delta^2}$ & $\boldsymbol{\Delta_U^2}$ & $\boldsymbol{\Delta_U}$ & $\boldsymbol{\Delta^2}$ & $\boldsymbol{\Delta_U^2}$ & $\boldsymbol{\Delta_U}$ & $\boldsymbol{\Delta^2}$ & $\boldsymbol{\Delta_U^2}$ & $\boldsymbol{\Delta_U}$ \\
    $(\mathrm{h~Mpc^{-1}})$  & $(\mathrm{mK^2})$   & $(\mathrm{mK^2})$   & $(\mathrm{mK})$    & $(\mathrm{mK^2})$   & $(\mathrm{mK^2})$   & $(\mathrm{mK})$    & $(\mathrm{mK^2})$   & $(\mathrm{mK^2})$   & $(\mathrm{mK})$ \\
    \midrule
    \textbf{0.135} & 5.12E+03 & 5.44E+03 & 73.78 & 8.73E+03 & 9.28E+03 & 96.34 & 1.43E+04 & 1.52E+04 & 123.25 \\
    \textbf{0.177} & 2.05E+04 & 2.13E+04 & 145.97 & 2.95E+04 & 3.06E+04 & 175.05 & 4.54E+04 & 4.71E+04 & 216.91 \\
    \textbf{0.248} & 4.05E+04 & 4.23E+04 & 205.55 & 6.36E+04 & 6.61E+04 & 257.19 & 1.07E+05 & 1.11E+05 & 333.57 \\
    \textbf{0.319} & 1.37E+05 & 1.44E+05 & 379.12 & 1.51E+06 & 1.55E+06 & 1246.86 & 2.33E+06 & 2.40E+06 & 1550.42 \\
    \textbf{0.389} & 2.10E+07 & 2.16E+07 & 4649.95 & 8.76E+07 & 9.02E+07 & 9498.26 & 1.34E+08 & 1.38E+08 & 11758.27 \\
    \textbf{0.460} & 1.04E+08 & 1.07E+08 & 10327.4 & 5.13E+07 & 5.29E+07 & 7273.08 & 8.05E+07 & 8.29E+07 & 9106.82 \\
    \textbf{0.531} & 4.93E+06 & 5.08E+06 & 2254.25 & 3.72E+05 & 3.95E+05 & 628.23 & 6.53E+05 & 6.84E+05 & 827.07 \\
    \textbf{0.602} & 3.58E+05 & 3.76E+05 & 612.94 & 6.54E+05 & 6.90E+05 & 830.54 & 1.33E+06 & 1.39E+06 & 1177.4 \\
    \textbf{0.672} & 1.65E+06 & 1.71E+06 & 1305.86 & 2.20E+06 & 2.28E+06 & 1511.57 & 3.43E+06 & 3.56E+06 & 1886.78 \\
    \textbf{0.743} & 1.72E+06 & 1.78E+06 & 1336.03 & 2.35E+06 & 2.45E+06 & 1564.64 & 3.26E+07 & 3.36E+07 & 5800.56 \\
    \textbf{0.814} & 2.06E+08 & 2.12E+08 & 14555.44 & 2.94E+08 & 3.03E+08 & 17404.05 & 1.28E+09 & 1.32E+09 & 36371.56 \\
    \textbf{0.885} & 7.52E+08 & 7.75E+08 & 27841.21 & 1.08E+09 & 1.12E+09 & 33406.26 & 5.76E+08 & 5.94E+08 & 24369.03 \\
    \textbf{0.956} & 2.77E+07 & 2.86E+07 & 5348.69 & 3.95E+07 & 4.07E+07 & 6382.98 & 2.80E+06 & 2.95E+06 & 1718.88 \\
    \textbf{1.026} & 9.88E+05 & 1.08E+06 & 1041.16 & 2.45E+06 & 2.59E+06 & 1607.81 & 5.22E+06 & 5.46E+06 & 2337.5 \\
    \bottomrule
    \multicolumn{10}{l}{$^{4}$} \\
    \end{tabular}
\caption{Measured power($\Delta^2$), 2$\sigma$ upper limit ($\Delta_U^2$) and the square root of the upper limit ($\Delta_U$) for three redshifts. The reported k values point to the bin centre.}
\label{table:limits}
\end{table*}

\subsection{Comparison with Previous Results}\label{sec:comp-with-previous}
The MWA high-band EoR1 data processing by RTS/CHIPS pipeline has only been previously studied by \citet{Trott2020}. In this section, we discuss the results and address the differences in the two analysis approaches, as summarised in tables \ref{tbl:comp-cath} and \ref{tab:limit_compare}.
\begin{table*}
\small
    \centering
    \begin{tabular}{lll}
     \hline
 Feature  &  This work  & \citet{Trott2020}\\
 \hline
 No. of Selected Observations & 459 & 600\\
 Observation Time & 2015/2014 & 2013-2016\\
 Pointings  & five central & zenith\\
 MWA Phase of operation & phase-I & phase-I/II\\
 Source catalogue & puma2020   & puma2017a/b \\
 Shapelet Modeling & new model based on phaseI/II image of Fornax & old model   \\
 IonoQA metric cut-off &  5  & 30\\
 Other Applied Metrics & auto-correlation & N/A \\
  & RMS trend & \\
  & Calibration Solution Profiles & \\
  \hline
    \end{tabular}
    \caption{Comparison of analysis approach of this work and \citet{Trott2020}}
    \label{tbl:comp-cath}
\end{table*}

\begin{table}
    \centering
    \begin{tabular}{llll}
    \hline
  &\multicolumn{3}{c}{$\boldsymbol{\Delta_U}(mK)$} at k=0.14 ($\mathrm{h~Mpc^{-1}}$)\\ 
   &z=6.5 &z=6.8 & z=7.1 \\
  \hline
  This work (459-observation) & 145.970 & 175.05 & 216.91 \\
  \citet{Trott2020} (600-observation) & 183.8  & 199.9& 305.0 \\
  \hline
    \end{tabular}
    \caption{Comparison of the upper limits between this work and \citet{Trott2020} at $k=0.14 ~\mathrm{h~Mpc^{-1}}$ which is the lowest common k.}   \label{tab:limit_compare}
\end{table}

\begin{figure}
    \centering
    \includegraphics[width=0.9\columnwidth]{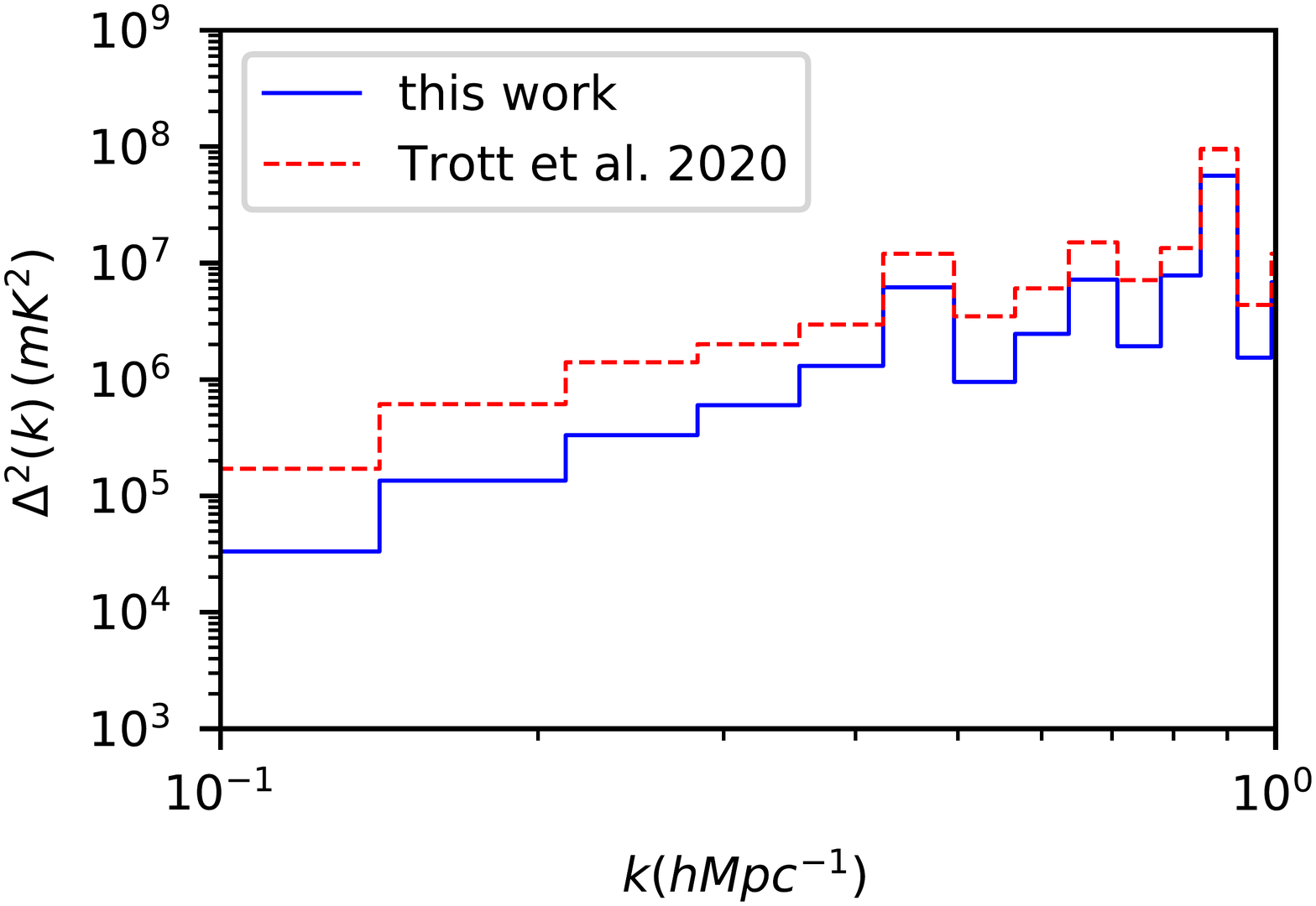}
    \caption{Comparison of 1D power spectra of this work (blue) and \citet{Trott2020} (red) for 20 observations.}
    \label{fig:comp_cath}
\end{figure}

As discussed in section \ref{sec:improvement}, the upgraded version of RTS and source catalogues have significantly improved data reduction and power measurement results of EoR1 data. The \citet{Trott2020} study also includes 2013 data which suffers from lack of dead dipole information. With more conservative data refinement strategies, a less systematic-contaminated power measurement has been obtained. In figure \ref{fig:comp_cath}, the 1D power spectra of a sample of 20 observations reduced with both versions of the pipelines are compared.

\subsection{Discussion}
Currently, the MWA EoR power spectrum measurements do not reach the expected thermal noise and hence are limited by systematics. In this work a 14-hr integration using an improved (RTS+CHIPS) analysis pipeline has been demonstrated the possibility of lowering this systematic floor. By deeper integration of observations and further mitigation of systematic errors, we will potentially be able to reach the noise-dominated level of signal power and ultimately detect the EoR signal. In this regime, the assurance of signal preservation will be  critical and requires a full end-to-end simulation which is beyond the scope of this work and is a part of the collaboration research program. However, in the case of this work in which the integration time is short and measurements are higher than the noise level, the possible signal loss does not largely interfere with our results. Nevertheless the lack of signal loss issue is (in)directly validated by different probes: As studied in the recent deep measurements of EoR upper limits with RTS+CHIPS, \citet{Trott2020} showed that there is no signal loss occurring in the CHIPS pipeline. It is also shown that there is no signal loss occurring in RTS+CHIPS due to imperfect telescope modelling. In the measurements by the the parallel MWA EoR analysis pipeline (FHD+ $\varepsilon$ppsilon), lack of signal loss is also validated via simulation\citep{Barry:2016,Barry2019,Li2019}. The consistent results from cross-checking the two independent parallel pipelines, assures the validity of signal loss test results for both pipelines \citep{Jacobs:2016,Beardsley2016,Barry2019,Li2019}.

\section{Conclusions}
This work has presented the analysis of Phase-I MWA EoR data targeted at the EoR1 field using the RTS/CHIPS pipeline. This data includes five central pointing observations of the field in 2015/2014. By applying more cautious data cleaning strategies, whilst improving the analysis pipeline and source modelling, we measure the upper limit of the power spectrum at three redshift bands centered at 6.5, 6.8 and 7.1, with the lowest measurement obtained at $z=6.5$ of $\Delta^2 \leq (73.78~\mathrm{mK})^2$ at $k=0.13~\mathrm{h Mpc^{-1}}$. The upper limits are the result of integrating $\sim14~\mathrm{hr}$ of data, which despite shorter integration time is 1.26 times lower (at $k=0.14 \mathrm{h Mpc^{-1}}$) than the previous EoR1 analysis results by \citet{Trott2020} which is a $\sim 19~\mathrm{hr}$ integration. This improvement confirms the necessity of improving the modelling of bright and extended sources with significant spectral structure such as Fornax-A to achieve more robust foreground mitigation, in addition to removing potentially contaminated data via visibility RMS metrics. Although the cuts we used are motivated by removing systematics, they can also lead to signal loss if the data is not systematically limited. In future, we need to expand our signal loss simulation to a more representative sample to show their effect on the cosmological signal.

We also studied the difference between two MWA EoR fields, EoR0 and EoR1, each characterised by different foregrounds which changes the sky temperature. EoR0 is more contaminated by galactic emission on large scales, while EoR1 includes the bright radio source of Fornax-A. As shown in the comparison power spectra of the two fields in figure \ref{fig:eor0-eor1-ps}, while EoR1 is generally more contaminated by sources than EoR0, the strong presence of galactic emission in EoR0 along the horizon border impacts the wide-field measurements at large scale. The EoR1 analysis in this work and \citet{Trott2020} include $\sim 14$ and $19~\mathrm{hour}$ integration, respectively. The improvements in the analysis tools warrant performing a deeper analysis similar to that carried out by \citet{Trott2020} for EoR0, using 100s of hours of data.

In this work, we explored various aspects of the data to reduce the final upper limit: 
\begin{itemize}
    \item  Since the 2013 data does not contain information on the dead dipoles which affects the modeled beam response, additional systematics are introduced into the power measurement of any dataset which includes 2013 data. Using 2015/2014 data in this work, the selected dataset results in power measurements which are improved relative to previous published limits, despite the shorter integration time.
    \item As auto-correlations convey the same signature of error contamination as cross-correlations, they are a powerful tool for cleaning the data. In the case of MWA, they are also computationally efficient as they can be implemented without the need for calibration.
    \item One of the effective approaches to mitigate systematics is to avoid all affected data once a hardware-related issue is recognised. As an example, the high-RMS data studied in this work appeared to occur for a finite time duration on the scale of months for all LSTs, indicating a hardware-related issue. By flagging such datasets, we can prevent the introduction of identified systematics into the power spectrum.
\end{itemize}
Each new EoR data analysis can further illuminate the strengths and weaknesses of the existing EoR datasets and indicate new directions for the next generation of experiments. With the current development of SKA-Low and other new telescopes, these experiments can advise more effective and robust pipelines. This work again confirms how developing more precise sky models and upgrading analysis algorithms equips us for more effective mitigation of the systematic errors. However, even though many improvements have been made, we selected just $\sim60\%$ of data as a relatively reliable dataset to obtain less contaminated measurements. The data loss was due to poor observational conditions, poor analysis outcomes or poor quality data. Particularly, large quantities of the poor quality data we excised seems to be related to hardware issues, specifically regarding a few tiles within each observation. Thus, along with the efforts toward more robust analysis algorithms and sky models, it is also essential to sustain routines in the hardware platform to lessen data failures, hence extracting the most out of observing and processing hours.

\section*{Acknowledgements}

The authors wish to thank the anonymous referee for providing constructive suggestions that greatly improved the paper. This research was supported by Australian Government Research Training Program Scholarship and the Australian Research Council Centre of Excellence for All Sky Astrophysics in 3 Dimensions (ASTRO 3D), through project number CE170100013.

Computational resources used were awarded under Astronomy Australia Ltd’s ASTAC merit allocation scheme on the OzSTAR national facility at Swinburne University of Technology. The OzSTAR program receives funding in part from the Astronomy National Collaborative Research Infrastructure Strategy (NCRIS) allocation provided by the Australian Government.

This work makes use of the Murchison Radio-astronomy Observatory, operated by CSIRO. We acknowledge the Wajarri Yamatji people as the traditional owners of the Observatory site.

\section*{Data Availability}
The data used in this work is publicly available in MWA portal: \url{https://asvo.mwatelescope.org/}.



\bibliographystyle{mnras}
\bibliography{mybib.bib} 




\appendix
\section{Comparison of Sky Temperature of EoR0 and EoR1 Fields}\label{sec:sky-temp}
We compare the sky temperature over EoR0 and EoR1 patches by means of both the Global Sky Model (GSM, \citet{deOliveira-Costa:2008}) and directly from our observational data.

\begin{figure}
    \centering
    \subfloat[GSM predicted sky temperature vs. LST at $180~\mathrm{MHz}$.]{
        \includegraphics[width=0.9\columnwidth]{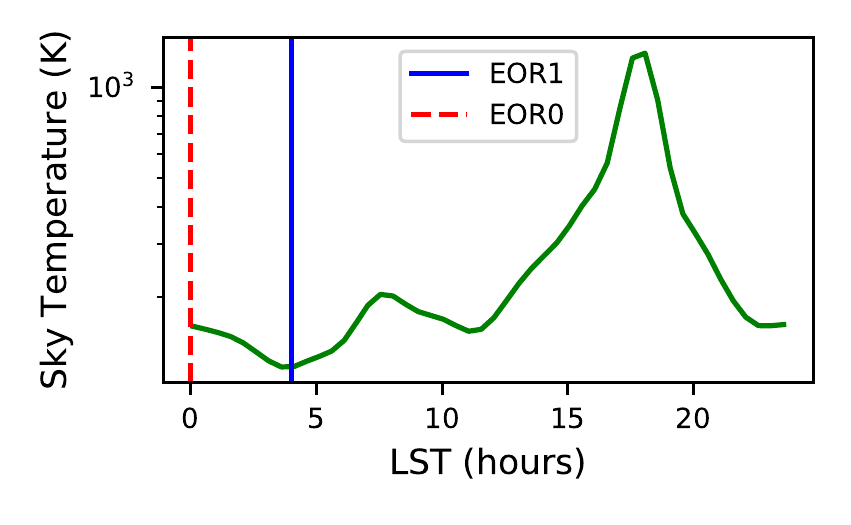}
        \label{fig:gsm_lst}
        }
        \\
    \subfloat[GSM predicted Sky temperature as a function of frequency for EoR0 and EoR1 fields.]{
        \includegraphics[width=0.9\columnwidth]{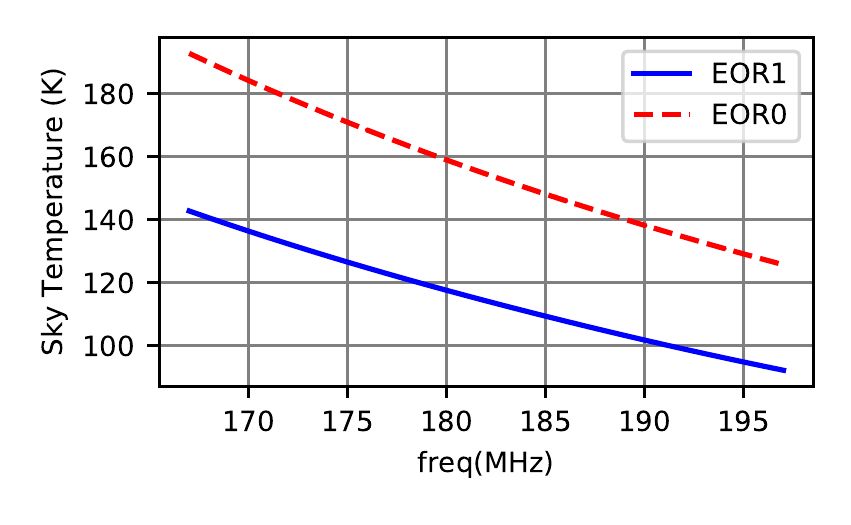}
        \label{fig:gsm-eor0-eor1}
        }
    \caption{GSM predicted sky temperature with MWA zenith-pointing beam, indicating EoR1 is a colder patch of sky comparing to EoR0.}
    \label{fig:gsm}
\end{figure}
In figure \ref{fig:gsm}, the GSM prediction of the sky temperature in the EoR0 and EoR1 fields is illustrated. The GSM data has been weighted by the zenith-pointing MWA beam. Figure \ref{fig:gsm_lst} shows the sky temperature as a function of Local Sidereal Time(LST) at a frequency of $180~\mathrm{MHz}$ which is the MWA high-band central frequency. EoR0 is observed at zenith at  $LST=0~\mathrm{hr}$ and EoR1 at $LST=4~\mathrm{hr}$. The plot indicates the lower temperature of EoR1($\sim 117~\mathrm{K}$) compared to EoR0($\sim 159~\mathrm{K}$) at $180~\mathrm{MHz}$, while also showing the peak at $LST=17~\mathrm{hr}$ which corresponds to observing the Galactic Centre at MWA zenith. Figure \ref{fig:gsm-eor0-eor1} also illustrates the trend of sky temperature over the whole MWA frequency band for EoR0 and EoR1, confirming a theoretically colder sky at EoR1.

\begin{figure}
    \centering
    \includegraphics[width=0.95\linewidth]{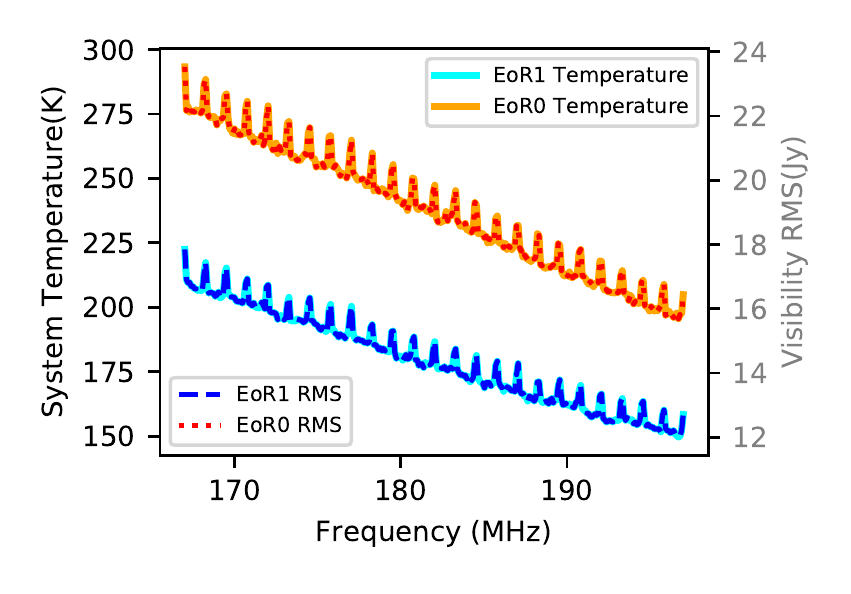}
    \caption{System temperature (scaled on the left vertical axis) and visibility noise RMS (scaled on the right vertical axis) for EoR0(in dotted orange and solid red, respectively) and EoR1 (in dashed blue and solid aqua, respectively). EoR1 visibilities show less visibility noise which agrees with lower sky temperature over this field.}
    \label{fig:rms-temp}
\end{figure}
To demonstrate these differences with our observational data, we take advantage of the temperature information embedded in the visibility noise RMS. Following \citet{Thompson:1999}, $T_{sys}$ can be obtained as:
\begin{equation}
    T_{sys} = \frac{A\sigma_{vis,p}\sqrt{df dt}}{k_B}
\end{equation}
in which $A$ is the effective area per tile, $\sigma_{vis,p}$ is the visibilities noise RMS of a single polarization $p$, as defined in section \ref{sec:comp_13_15}, $df$ and $dt$ are the frequency and time resolution of the observations and $k_B$ is the Boltzmann constant. The exact same analysis was performed on the EoR1 field and the EoR0 field, and thus the main difference in the visibility RMS should be related to the system temperature. In figure \ref{fig:rms-temp}, the RMS and corresponding system temperatures of both EoR1 and EoR0 are compared, giving a system temperature of $\sim 240~\mathrm{K}$ for EoR0 and $\sim 180\mathrm{K}$ for EoR1 at $180~\mathrm{MHz}$. The observationally measured system temperature of a radio interferometer, $T_{sys}$, is the sum of the sky temperature and the receiver temperature. Thus, receiver temperatures of $\sim 63~\mathrm{K}$ and $\sim 81~\mathrm{K}$ for EoR1 and EoR0 respectively are deduced. The estimated receiver temperatures are broadly consistent with the reported temperature of $50~\mathrm{K}$ at $150~\mathrm{MHz}$ by \citet{Tingay:2013} but do indicate some unidentified systematic effects which also differ between the fields. This discrepancy might be explained by the possible differences of the related components between the two fields; e.g. the basis of GSM model is the Galactic diffuse emission while not including the bright extragalactic sources such as Fornax-A (which is the dominant emission in case of EoR1). It should also be considered that the calculations of system temperature of the two fields are based on two typical individual observations which might not perfectly represent the actual system temperature of their field. The results are summarized in table \ref{tab:T_sky}. 
\begin{table}
    \centering
    \begin{tabular}{lcc}
    \toprule
    & GSM predicted & Observationally Driven \\
    Field & Sky Temperature ($\mathrm{K}$) & System Temperature ($\mathrm{K}$) at $180 \mathrm{MHz}$.\\
    \midrule
    EoR0 & 159 & 240  \\
    EoR1 & 117 & 180  \\
    \bottomrule
    \end{tabular}
    \caption{Sky/System Temperature at MWA EoR fields.}
    \label{tab:T_sky}
\end{table}

\section{Recovering Simulated EoR Signal with RTS/CHIPS Pipeline}\label{app:sim}
In order to validate our normalisation and provide a measure of confidence in our limits, we performed an end-to-end simulation to generate power spectra from a known input model. We chose a fully neutral IGM power spectrum model of the EoR from \citet{Furlanetto:2006} as our input. In order to keep the simulation simple, we generated a measurement uv-plane with Gaussian noise which emulates the desired power spectrum. Then, through a process called degridding, we integrated the product of the uv-plane and the uv representation of the instrumental beam at every baseline location. This creates instrumental visibilities which are then analysed by the RTS/CHIPS pipeline to generate an output power spectrum. Calibration was not performed in this normalisation validation. 

\begin{figure}
    \centering
    \includegraphics[width=0.9\columnwidth]{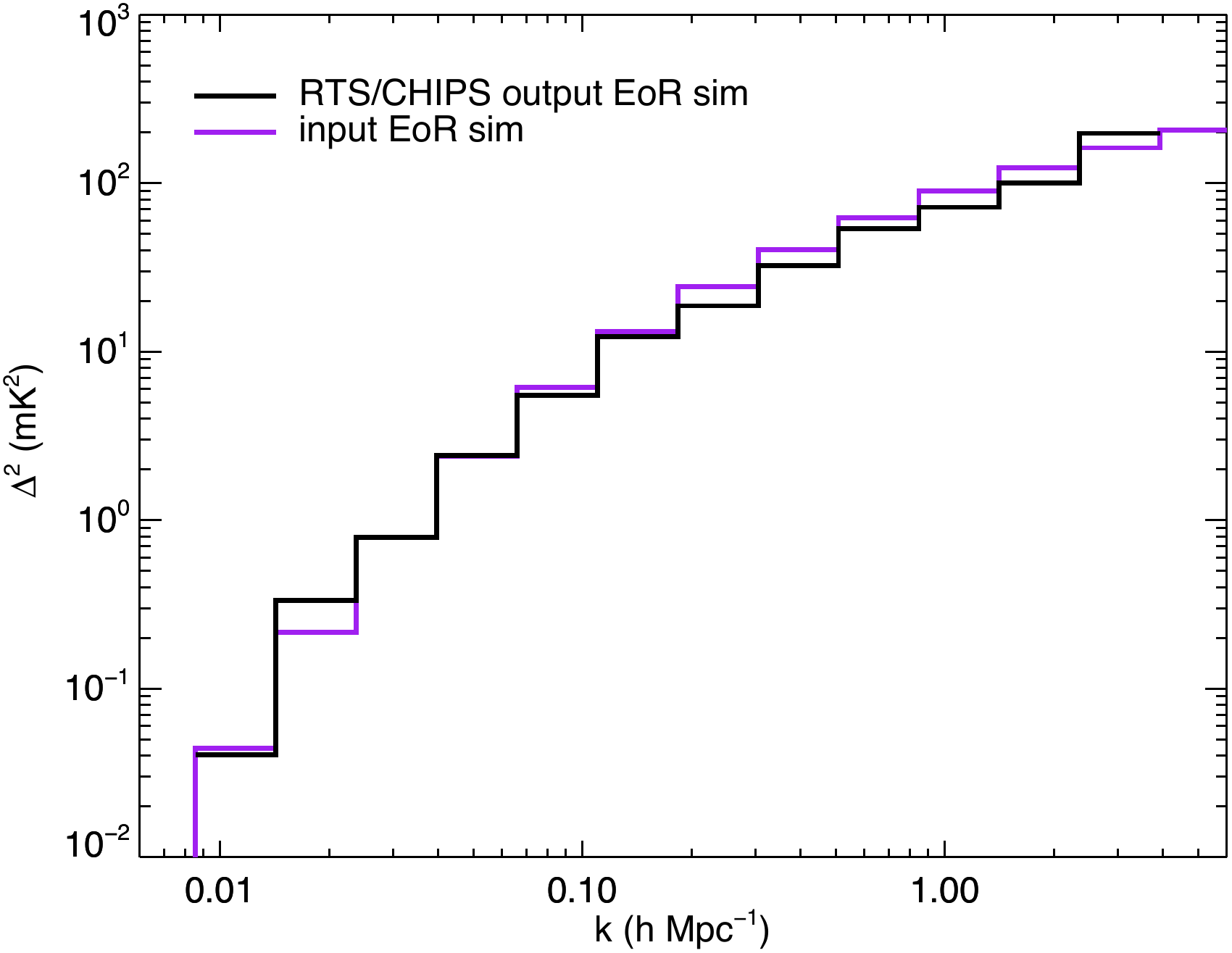}
    \caption{Input power spectrum of a theoretical EoR signal (purple) and the recovered output signal through the RTS/CHIPS pipeline (black). The normalisation of the output power spectrum is consistent with the input, thus giving confidence to the normalisation of our EoR upper limits in the absence of calibration.}
    \label{fig:sim}
\end{figure}

Figure~\ref{fig:sim} shows both the input power spectrum from \citet{Furlanetto:2006} (purple) and the output power spectrum from simulated visibilities analysed by RTS and CHIPS (black). Given the agreement, we are confident in our overall normalisation of our power spectrum limits. While there are some differences, this can be investigated in the future using image realisations of the EoR, foregrounds, and various calibration schemes.

\bsp	
\label{lastpage}
\end{document}